\documentclass[a4paper,fleqn,usenatbib]{mnras}
\usepackage{ae,aecompl}
\usepackage{graphicx}	
\usepackage{amsmath}	
\usepackage{amssymb}	
\usepackage{amsbsy}
\usepackage{float}
\usepackage{color}
\usepackage{txfonts}
\usepackage{times}

\usepackage{morefloats}
\usepackage{multirow}

\usepackage{mathrsfs,bm}
\newcommand{\bcdot}{\ensuremath{%
  \mathchoice%
   {\mskip\thinmuskip\lower0.2ex\hbox{\scalebox{1.5}{$\cdot$}}\mskip\thinmuskip}}%
   {\mskip\thinmuskip\lower0.2ex\hbox{\scalebox{1.5}{$\cdot$}}\mskip\thinmuskip}%
   {\lower0.3ex\hbox{\scalebox{1.2}{$\cdot$}}}%
   {\lower0.3ex\hbox{\scalebox{1.2}{$\cdot$}}}%
}

\title[Matter and galaxy clustering in IllustrisTNG]
{First results from the IllustrisTNG simulations: matter and galaxy clustering}

\author[V.~Springel et al.] {Volker Springel$^{1,2}$, R\"udiger~Pakmor$^1$, Annalisa Pillepich$^{3}$,  Rainer Weinberger$^{1}$, \newauthor
Dylan Nelson$^{4}$, Lars Hernquist$^{5}$, Mark Vogelsberger$^{6}$, Shy Genel$^{7,8}$,  Paul Torrey$^{6}$, \newauthor
Federico Marinacci$^{6}$, Jill Naiman$^{5}$
  \vspace*{0.2cm}  \\
  $^1$Heidelberger Institut f\"{u}r Theoretische Studien,
  Schloss-Wolfsbrunnenweg 35, 69118 Heidelberg, Germany\\
  $^2$Zentrum f\"ur Astronomie der Universit\"at Heidelberg, ARI, M\"onchhofstrasse
12-14, 69120 Heidelberg, Germany\\
  $^3$Max-Planck-Institut f\"ur Astronomie, K\"onigstuhl 17, 69117 Heidelberg, Germany\\
  $^4$Max-Planck-Institut f\"ur Astrophysik, Karl-Schwarzschild-Str. 1, D-85748, Garching, Germany\\
  $^5$Harvard-Smithsonian Center for Astrophysics, 60 Garden Street, Cambridge, MA 02138, USA\\
  $^6$Department of Physics, Kavli Institute for Astrophysics and Space Research, MIT, Cambridge, MA 02139, USA\\
$^7$Center for Computational Astrophysics, Flatiron Institute, 162 Fifth Avenue, New York, NY 10010, USA\\
$^8$Columbia Astrophysics Laboratory, Columbia University, 550 West 120th Street, New York, NY 10027, USA
}
  
\pubyear{2017}

\renewcommand{\vec}[1]{ {\bf #1} }

\begin{document}

\label{firstpage}
\pagerange{\pageref{firstpage}--\pageref{lastpage}}

\maketitle

\begin{abstract}
  Hydrodynamical simulations of galaxy formation have now reached
  sufficient volume to make precision predictions for clustering on
  cosmologically relevant scales. Here we use our new IllustrisTNG
  simulations to study the non-linear correlation functions and power
  spectra of baryons, dark matter, galaxies and haloes over an
  exceptionally large range of scales.  We find that baryonic effects
  increase the clustering of dark matter on small scales and damp the
  total matter power spectrum on scales up to
  $k \sim 10\,h\,{\rm Mpc}^{-1}$ by $20\%$.  The non-linear two-point
  correlation function of the stellar mass is close to a power-law
  over a wide range of scales and approximately invariant in time from
  very high redshift to the present. The two-point correlation
  function of the simulated galaxies agrees well with SDSS at its mean
  redshift $z \simeq 0.1$, both as a function of stellar mass and when
  split according to galaxy colour, apart from a mild excess in the
  clustering of red galaxies in the stellar mass range
  $10^9-10^{10}\,h^{-2}{\rm M}_\odot$. Given this agreement, the TNG
  simulations can make valuable theoretical predictions for the
  clustering bias of different galaxy samples. We find that the
  clustering length of the galaxy auto-correlation function depends
  strongly on stellar mass and redshift. Its power-law slope $\gamma$
  is nearly invariant with stellar mass, but declines from
  $\gamma \sim 1.8$ at redshift $z=0$ to $\gamma \sim 1.6$ at redshift
  $z\sim 1$, beyond which the slope steepens again.  We detect
  significant scale-dependencies in the bias of different
  observational tracers of large-scale structure, extending well into
  the range of the baryonic acoustic oscillations and causing nominal
  (yet fortunately correctable) shifts of the acoustic peaks of around
  $\sim 5\%$.
\end{abstract}


\section{Introduction}

Ever since the discovery of cosmic large-scale structure
\citep{Geller1989, Bond1996}, the clustering of galaxies has been
recognised as one of the most important observational constraints in
cosmology \citep[e.g.][]{Tegmark2004, Sanchez2006}. Galaxy redshift
surveys have found early on that the two-point autocorrelation
functions of different types of galaxies are close to power-laws at
low-redshift \citep{Davis1983}, and that they evolve little with time
over the range where observational constraints are available, in stark
contrast to the predicted rapid change of the autocorrelation function
of the underlying mass distribution in cold dark matter cosmologies
\citep{Davis1985, Jenkins1998}. Galaxies are thus at best a biased
tracer of the matter fields \citep{Kaiser1984, Davis1985,
  White1987}. In general, this bias relative to the mass distribution
is much larger at high redshift than in the present epoch
\citep[e.g.][]{Springel2006}, and it exhibits an interesting
scale-dependence that reconciles the very different shapes of the
matter and galaxy correlation functions.

The fact that there is a significant galaxy bias on large scales can
be readily understood from the expected clustering signal of dark matter
haloes when they are associated with the peaks of Gaussian random
fields \citep{Bardeen1986}. Insightful analytic models for the bias of
dark matter haloes as a function of their mass exist \citep{Mo1996,
  Sheth1999}, and when combined with a prescription for how galaxies
populate the haloes, approximate forecasts for the galaxy bias can be
obtained. However, the quantitative accuracy of these predictions is
difficult to assess without detailed simulation models. In addition,
on intermediate and small cosmological scales, the bias becomes scale
dependent, something not readily accessible in simple theories of
galaxy bias \citep[see][for an extensive review of the theory of
bias]{Desjacques2016}. However, a precise understanding of galaxy bias
is necessary in order to make optimum use of forthcoming cosmological
surveys (e.g.~DES, eBOSS, DESI, or EUCLID), in particular those that
target dark energy. Simply discarding all data on scales that may be
polluted by non-linear bias may severely degrade the constraining
power of these surveys.

To make full use of the observational data and properly understand
potential systematic effects due to galaxy bias, it is imperative to
have self-consistent physical models of galaxy formation that link
galaxy properties directly to the evolving matter fields. Such models
encode our best theoretical understanding for how galaxies may have
formed and can also properly account for second-order effects such as
assembly bias \citep{Gao2005, Yang2006, Wechsler2006, Wang2013,
  Zentner2014} or galactic conformity \citep{Kauffmann2015, Bray2016}.

Semi-analytic models of galaxy formation coupled to subhalo merging
trees extracted from dark matter-only simulations
\citep{Kauffmann1999, Springel2001, Springel2005}, have for a long
time been one of the most successful approaches to predict the
large-scale clustering of galaxies \citep[e.g.][]{Guo2011}. Here, a
large volume can be reached, and the parameterisation of galaxy
formation physics used in these models has achieved a high degree of
sophistication, matching a large variety of observational data, both
at the present epoch and at high redshift
\citep[e.g.][]{Kauffmann1993, DeLucia2006, Somerville2008, Benson2012,
  Henriques2015, Clay2015, Croton2016, Lacey2016, Cattaneo2017}.

A simpler alternative are subhalo abundance matching (SHAM) models
\citep{Moster2010, Behroozi2010, Guo2010, Masaki2013, Campbell2017}, or (still
simpler) halo occupation distribution (HOD) approaches
\citep{Peacock2000, Berlind2002}. While they lack a clear physical
basis and are largely empirically based, they are very popular as a
simple means to model large amounts of galaxy survey data. They do not
properly capture effects such as assembly bias, but efforts have been
made to outfit these empirical techniques with additional
environmental dependencies to address this deficiency
\citep{Hearin2016}. We note that some of the most recent empiric
models for galaxy formation \citep{Zu2015, vanDaalen2016, Moster2017} actually
employ galaxy clustering data as an {\em input constraint}, thereby
limiting their ability to predict large-scale structure observables.

Explicit comparisons between different semi-analytic models and HOD
approaches have shown that they can differ significantly in their
clustering predictions due to the different treatments of orphans and
satellite galaxies \citep{Pujo2017}. Similarly,
\citet{Chaves-Montero2016} have measured the two-point correlation
function of galaxies in the EAGLE simulation \citep{Schaye2005Eagle}
in various mass bins, finding systematic deviations to SHAM models for
the same simulation.  By construction, neither the semi-analytic
models nor the empirical SHAM/HOD approaches offer detailed
predictions for the clustering of the baryonic matter, nor can they
account for the back-reaction of baryons on the clustering of the dark
matter, which is associated with strong feedback effects. This
omission of an explicit modeling of hydrodynamical processes thus adds
significant theoretical uncertainty in these models
\citep{Guo2016}. Hydrodynamical simulations are much more constraining
and powerful in this respect, even though they also still need to
invoke empirical input to parameterise uncertain feedback physics on
small, unresolved scales.

Early efforts to use hydrodynamical simulations of galaxy formation to
predict galaxy clustering \citep{Katz1999, Weinberg2004, Nuza2010}
were severely challenged by the small size of the feasible
cosmological volumes at the time, the comparatively low numerical
resolution that could be achieved, and the still limited understanding
of the feedback physics. Progress has slowly been made over the years
on all of these fronts, but only the advent of a new generation of
hydrodynamic cosmological simulations over the last couple of years
has made this approach a serious competitor to semi-analytic and
empirical galaxy formation models. Projects such as Illustris
\citep{Vogelsberger2014, Genel2014}, EAGLE \citep{Schaye2005Eagle},
MassiveBlack-II \citep{Kandai2015}, HorizonAGN \citep{Dubois2016} and
Mageneticum \citep{Dolag2016} have succeeded in predicting galaxy
populations in reasonable agreement with observational constraints,
throughout quite large cosmological volumes, allowing in principle
realistic clustering predictions.  For example, \citet{DeGraf2017}
have studied the clustering of active galactic nuclei in Illustris,
including the bias of the black hole population relative to the dark
matter, and \citet{Crain2017} considered the clustering of atomic
hydrogen sources in EAGLE.

\begin{table*}
\begin{tabular}{llccrrrccc}
\hline
Series & Run &   \multicolumn{2}{c}{Boxsize}  
& $N_{\rm gas}$  & $N_{\rm dm}$ & $N_{\rm tracer}$
  & $m_{\rm b}$ & $m_{\rm dm}$ & $\epsilon$ \\
&        &   $[h^{-1}{\rm Mpc}]$  & $[{\rm Mpc}]$  & & &  
 & $[h^{-1}{\rm M}_\odot]$ & $[h^{-1}{\rm M}_\odot]$ 
&  $[h^{-1}{\rm kpc}]$ \\
\hline 
{\bf TNG300} &  TNG300(-1)  & 205  & 302.6 &
 $2500^3$ & $2500^3$ & $2500^3$
& $7.44\times 10^6$ & $3.98\times 10^7$ & 1.0 \\
                     &  TNG300-2  & 205  & 302.6 &
 $1250^3$ & $1250^3$ & $1250^3$
& $5.95\times 10^7$ & $3.19\times 10^8$ & 2.0 \\
                &  TNG300-3  & 205  & 302.6 &
 $625^3$ & $625^3$ & $625^3$
& $4.76\times 10^8$ & $2.55\times 10^9$ & 4.0 \\
     &  TNG300-DM(-1)  & 205  & 302.6 &
   & $2500^3$ &  
&   & $4.73\times 10^7$ & 1.0 \\
     &  TNG300-DM-2  & 205  & 302.6 &
   & $1250^3$ &  
&   & $3.78\times 10^8$ & 2.0 \\
     &  TNG300-DM-3  & 205  & 302.6 &
   & $625^3$ &  
&   & $3.03\times 10^9$ & 4.0 \\
\hline
{\bf TNG100} &  TNG100(-1)  & 75  & 110.7 &
 $1820^3$ & $1820^3$ & $2 \times 1820^3$
& $9.44\times 10^5$ & $5.06\times 10^6$ & 0.5 \\
                     &  TNG100-2  & 75  & 110.7 &
 $910^3$ & $910^3$ & $2 \times 910^3$
& $7.55\times 10^6$ & $4.04\times 10^7$ & 1.0 \\
                &  TNG100-3  & 75  & 110.7 &
 $455^3$ & $455^3$ & $2 \times 455^3$
& $6.04\times 10^7$ & $3.24\times 10^8$ & 2.0 \\
     &  TNG100-DM(-1)  & 75  & 110.7 &
   & $1820^3$ &  
&   & $6.00\times 10^6$ & 0.5 \\
     &  TNG100-DM-2  & 75  & 110.7 &
   & $910^3$ &  
&   & $4.80\times 10^7$ & 1.0 \\
     &  TNG100-DM-3  & 75  & 110.7 &
   & $455^3$ &  
&   & $3.84\times 10^8$ & 2.0 \\
\hline
\end{tabular}
\caption{Basic numerical parameters of the two primary runs of the
  IllustrisTNG simulation suite that are used here. We have carried 
  out simulations in three different periodic box sizes, 
  roughly of size $300$, $100$, and $50\,{\rm Mpc}$ on a side,
  as reflected in the individual simulation names, and analyse the two larger boxes 
  in this paper. For each box size, we have run different
  numerical resolutions spaced by a factor of 8 in mass
  resolution. The runs with gaseous cells are all full physics
  simulations which also included tracer particles, 
  and for each of them, we have carried out a
  corresponding dark matter only simulation as well. The values quoted
  for $m_{\rm b}$ and $m_{\rm dm}$ give the baryonic 
  (gas cells and star particles) and dark matter mass resolutions,
  respectively. The
  gravitational softening lengths $\epsilon$ refer to the maximum
  physical softening length of dark matter and star particles. The
  softening of gaseous cells is tied to their radius and allowed to
  fall below this value.   \label{tabsims}}
\end{table*}

Still, the enormous cost of these calculations makes it difficult to
simultaneously reach high enough spatial resolution to adequately
track galaxy formation and to have, at the same time, large enough
volume to study galaxy clustering. As a result, the analysis of galaxy
clustering in hydrodynamical simulations has been typically restricted
to quite small scales, as in \citet{Artale2016} for EAGLE, or
\citet{Kandai2015} for MassiveBlack-II.

In this work, we aim to make a significant step forward in this
regard. Our new TNG300 simulation employs a refined galaxy formation
model and improved numerical treatments, and it expands the volume by
a factor of 20 with respect to Illustris and EAGLE, while thanks to
the use of 31.125 billion resolution elements it still has
sufficiently high resolution to track galaxy formation substantially
below $L_\star$.  While our mass resolution in this large-volume
simulation is 8 times lower than in Illustris, it is still more than
20 times better than, for example, in the Millennium
simulation. Through our other new simulation, TNG100, carried out in a
smaller box and with higher mass resolution (equivalent to Illustris),
we can furthermore explicitly check for numerical convergence and
robustness on the scales that are represented in this smaller,
Illustris-like box.

The new TNG simulation model allows us to make interesting predictions
for the clustering of matter, including the gaseous, stellar and
supermassive black hole components, far into the non-linear regime and
over a wider range of scales than previously explored with
hydrodynamical simulations. We can directly use the simulated galaxies
to examine the relation of their clustering signal to the underlying
matter distribution, an analysis that is largely free of any
additional modelling assumptions. Finally, we can investigate how the
clustering of haloes and matter is impacted by baryonic effects. Given
that models for halo clustering are often used in analytical and
semi-analytical works in cosmology, even subtle effects here could be
quantitatively very important.

Clustering is most commonly studied either in real space through the
autocorrelation function or in Fourier space by means of the power
spectrum. While both viewpoints are Fourier transforms of each other
and are thus theoretically equivalent, in practice they entail
different measuring challenges and systematic effects. Hence they both
are useful complementary ways of analysing data and comparing to
theory. We will therefore repeatedly give results both for the
autocorrelation function and the power spectrum, hoping that this
improves the utility of our findings for the community.

This paper is structured as follows. In Section~\ref{sec_methods}, we
introduce our simulation methodology and discuss technical aspects of
our analysis. In Section~\ref{sec_matter}, we present results for the
clustering of different matter components, while in
Section~\ref{sec_galaxies} we extend this to different galaxy samples.
In Section~\ref{sec_haloes}, we consider the clustering of haloes in
our simulations and their linear bias on large scales.  In
Section~\ref{sec_bias}, we then turn to the bias of galaxies and its
dependence on stellar mass, redshift, and scale.  Finally, we discuss
our results and summarise our conclusions in
Section~\ref{sec_conclusions}.

\section{Methods}   \label{sec_methods}

\subsection{Simulation set}

{\it The Next Generation Illustris
Simulations}\footnote{\url{http://www.tng-project.org}} (IllustrisTNG)
studied here are an ambitious suite of new hydrodynamical simulations
of galaxy formation in large cosmological volumes. They are carried
out with the moving-mesh code {\small AREPO} \citep{Springel2010} and
use an updated galaxy formation model described in detail in
\citet{Weinberger2017} and \citet{Pillepich2017}. The most important
physics changes with respect to our previous Illustris simulation
physics model \citep{Vogelsberger2013} are an updated kinetic AGN
feedback model for the low accretion state \citep{Weinberger2017}, an
improved parameterisation of galactic winds \citep{Pillepich2017}, and
the inclusion of magnetic fields based on ideal magneto-hydrodynamics
\citep{Pakmor2011, Pakmor2013, Pakmor2014}. There have also been
numerous technical advances in the underlying simulation code, such as
improvements in the convergence rate of the hydrodynamical scheme
\citep{Pakmor2016} and the use of a more flexible hierarchical time
integration for gravitational interactions (Springel et al., in prep).

For the sake of brevity, we refer to the above publications and
references therein for a full description of the galaxy formation
model and the code, and tests carried out for it. We emphasise that
all model parameters of the IllustrisTNG runs have been kept exactly
the same as in our default model described in \citet{Pillepich2017},
and also no adjustments of these parameters are made for different
numerical mass resolutions, except for the gravitational softening
lengths and a sub-linear modification of the number of neighbouring
cells used in the black hole model. 

For IllustrisTNG, we have carried out simulations with three different
box sizes. TNG300 has a periodic box
$L=205\,h^{-1}{\rm Mpc} = 302.6\, {\rm Mpc}\sim 300\, {\rm Mpc}$ on a
side and a particle/cell number of $2\times 2500^3$ at the highest
resolution, which translates to a baryonic mass resolution of
$7.44\times 10^{6}\,h^{-1} {\rm M}_\odot$ and a dark matter particle
mass of $3.98\times 10^{7}\,h^{-1} {\rm M}_\odot$.  The simulation
series TNG100 has a box of intermediate size,
$L=75\,h^{-1}{\rm Mpc} = 110.7\, {\rm Mpc} \sim 100\, {\rm Mpc}$, and
uses a particle/cell number of $2\times 1820^3$ at its highest
resolution, the same as the Illustris simulation. Finally, TNG50 has a
small box with
$L=35\,h^{-1}{\rm Mpc} = 51.7\, {\rm Mpc} \sim 50\, {\rm Mpc}$ and up
to $2\times 2160^3$ resolution elements, pushing the baryonic mass
resolution down to $5.74 \times 10^{4}\,h^{-1} {\rm M}_\odot$. This
latter simulation is still in progress and is not analysed in this
paper. The gravitational softening lengths for dark matter and stars
in TNG300, TNG100, and TNG50 are $1.0\,h^{-1}{\rm kpc}$,
$0.5\,h^{-1}{\rm kpc}$, and $0.2\,h^{-1}{\rm kpc}$, respectively. The
softening of the mesh cells is adaptive and tied to their radii.

Besides carrying out these primary simulation boxes with full physics
at a nominally highest resolution, we have also run lower resolution
versions for each, which can be used to study numerical
convergence. We refer to them with an additional resolution
number. For example, `TNG300-1' is our highest resolution level,
`TNG300-2' has 8 times fewer resolution elements and two times worse
spatial resolution, while `TNG300-3' degrades the mass resolution by
another factor of 8 and the spatial resolution by a further factor of
2 with respect to `TNG300-2'. In addition, we have computed dark
matter only counterparts for all of these
simulations. Table~\ref{tabsims} gives an overview of the most
important numerical parameters of the simulation set analysed here.

The cosmology has been chosen in accordance with recent Planck
constraints \citep{PlanckCollab2016}\footnote{This is a change
  relative to our older Illustris project, which had been based on
  WMAP-9 measurements.}, and is given by
$\Omega_{\rm m} =\Omega_{\rm dm} + \Omega_{\rm b} = 0.3089$,
$\Omega_{\rm b} = 0.0486$, $\Omega_\Lambda=0.6911$, and Hubble
constant $H_0 = 100\,h\, {\rm km\, s^{-1}Mpc^{-1}}$ with $h=0.6774$.
The initial conditions were prescribed at $z=127$ using a linear
theory power spectrum computed for a normalisation $\sigma_8 = 0.8159$
and spectral index $n_s=0.9667$. When we compare to linear theory we
use this input spectrum, evolved to the corresponding redshift with
the linear growth factor.  The majority of the literature results on
clustering are expressed in units that retain a dependence on the
Hubble constant through
$h= H_0 / (100\,{\rm km\, s^{-1}\, Mpc^{-1})}$, with length units
given in $h^{-1}\,{\rm Mpc}$ and stellar mass units given in
$h^{-2}{\rm M}_\odot$, a convention we also retain here for the sake
of a simpler comparison. Note that the $h$-dependence of the stellar
mass unit originates in the conversion from apparent to absolute
magnitudes, whereas the natural theoretical mass unit, for example for
dark matter haloes, is $h^{-1}{\rm M}_\odot$.

This paper is one of five introductory studies of IllustrisTNG, each
concerned with a different scientific analysis topic enabled by the
simulations. The present work focuses on the galaxy and matter
clustering over a wide dynamic range. The other companion papers study
the colour-bimodality of galaxies \citep{Nelson2017TNG}, the properties
of the predicted magnetic fields \citep{Marinacci2017TNG}, the stellar
mass content of massive groups and clusters of galaxies
\citet{Pillepich2017TNG}, and the chemical enrichment of the elements magnesium and
europium \citep{Naiman2017TNG}.

\subsection{Power spectrum measurement}

The Fourier modes of the density contrast field for a set of $N$
points of mass $m_i$ in a periodic box of size $L$ can be defined as
\begin{equation}
\delta_{\vec{k}} = \frac{1}{M} \sum_{i} m_i \exp(i\,\vec{k}\cdot\vec{x}_i),
\end{equation}
where $M = \sum_i m_i$ is the total mass. The periodicity restricts
the available Fourier modes to integer multiples of $2\pi / L$ in each
dimension. Following the convention of
\citet{Peebles1980}, the power spectrum can then be defined as the
mean expected power per mode,
\begin{equation}
\hat P(k)  = \left< |\delta_{k}|^2\right>,
\end{equation}
which can be estimated through averaging the power of all modes
$\vec{k}$ with a length around a prescribed value of $k$. The power
spectrum may also be expressed in dimensionless form through
\begin{equation}
\Delta^2(k) = 4\pi k^3 P(k) / (2 \pi)^3,
\end{equation}
where now
\begin{equation}
 \Delta^2(k) = \frac{{\rm d}\sigma^2}{{\rm d} \ln k}
\end{equation}
gives the variance of the density field per unit $\ln k$.

To obtain the power spectra of different matter components or
galaxy/halo samples we use fast Fourier transforms (FFT) similar to
the methods employed in the TreePM gravity solver of the {\small
  AREPO} code.  To this end, the mass points are assigned with
cloud-in-cell (CIC) assignment to a uniform Cartesian mesh, thereby
obtaining a discrete representation of the density fluctuation field.
Upon Fourier transforming the density field, we obtain
$\delta_{\vec{k}}$ in Fourier space, which we deconvolve with the
smoothing effects of the kernel CIC assignment window. We then measure
the mean power per mode in a set of logarithmically spaced spherical
shells in $k$-space.

As is well known, estimating the power spectrum from a finite set of
random tracers in this way is affected by discreteness effects
\citep[e.g.][]{Colombi2009}. In particular, the power spectrum of a random
uniform distribution of points does not vanish, instead one obtains
so-called shot-noise power. For variable particle masses (as we have
here, especially for the black hole particles, and to a smaller extent
also for the stellar particles and gaseous cells), the shot noise
power is given by
\begin{equation}
P_{\rm shot} = L^3/N_{\rm eff},
\end{equation}
where $N_{\rm eff}$ can be viewed as an effective number of tracers,
given by
\begin{equation}
N_{\rm eff} = \frac{M^2}{\left<m^2\right>}.
\end{equation}
Here $M$ is the total mass of the tracers, and
$\left<m^2\right> = (\sum_i m_i^2)/N$ is the mean squared mass of the
individual tracers, with $N$ being their total number. For equal mass
tracers, $N_{\rm eff} = N$. If the tracer mass is dominated by a small
number of heavy particles, one can have $N_{\rm eff} \ll N$.

We typically estimate the power spectrum $P(k)$ of the underlying
density field by subtracting the shot-noise power from our raw estimate,
i.e.~we use
\begin{equation}
P(k) = \hat P(k) - P_{\rm shot}.
\end{equation} 
On small scales, this is fully adequate for the non-linearly clustered
dark matter and the stars, which represent Poisson samples of the
underlying density field. However, we note that dark matter haloes
have a finite size with some exclusion zone around them, such that the
shot noise correction for the halo power spectrum is only
approximately correct \citep[see, e.g.,][]{Smith2007}. Similarly, at
high-redshift, low density regions are still in the linear regime and
feature a relatively `cold' and ordered dark matter particle
distribution where the sampling is sub-Poissonian, so here the
shot-noise correction is generally too large. A small effect of this
kind is also present on small scales for the pressurised gas, leading
to a more regular point distribution than for a Poisson process.

We typically use base grids of size up to $4096^3$ for measuring the
power spectrum. Close to the Nyquist frequency
$k_{\rm Nyq} = \pi N_{\rm mesh} / {L}$ of the FFT mesh, aliasing
effects can create spurious amounts of excess power, therefore we only
consider $k < k_{\rm Nyq}/8$ as reliably measured. To fully measure
the power spectrum with a single mesh up to the highest resolved
spatial frequencies, $k_{\rm max} \simeq 2 \pi/\epsilon$, where
$\epsilon$ is the gravitational softening length, we therefore would
need a mesh of size $N_{\rm mesh}\simeq 10^5$, which is infeasible.
To extend the dynamic range, we therefore employ the `self-folding'
trick described in \citet{Jenkins1998} and compute power spectrum
measurements on smaller scales by mapping the box on top of itself
using a power-of-two subdivision $f_{\rm fold}$ of the full box.
Effectively, this imposes periodicity of the box on a smaller size
$L/f_{\rm fold}$, and a subsequent measurement of the power spectrum
determines only every $f_{\rm fold}^3$-th mode of the full
box. Because the number of modes on small scales is large, this still
allows a faithful measurement of the mean power per mode.

In order to cover the full dynamic range accessible in TNG300, we
actually apply the folding trick twice for a $4096^3$ mesh, once with
a folding factor $f_{\rm fold}=16$ and once with $f_{\rm
  fold}=16^2$. Even when staying below the Nyquist frequency by a
conservative factor of 8, this then gives an effective dynamic range
of $\sim 130,000$ between the largest and smallest scales that are
measured accurately -- enough for TNG300. As an alternative, it would
also be possible to use smaller FFTs and apply the folding trick more
frequently \citep{Colombi2009}.

\subsection{Correlation function measurements}

We measure the two-point correlation function of a point set in real
space using the classic definition
\begin{equation}
\xi(r) = \frac{\left< N_{\rm pairs}\right>}{N_{\rm mean}} - 1,
\end{equation} 
where $N_{\rm pairs}$ is the average number of other points found
around one of the points in a narrow spherical shell of radius $r$,
and $N_{\rm mean}$ is the mean number of points expected in the shell
for a uniform distribution of the points. If the particles have
variable mass, the points found in the shell are weighted by their
mass, $N_{\rm mean}$ is replaced by the mean mass in the shell, and
the contributions to the $\xi(r)$ estimate from each selected point
are weighted with the central point's mass. When measured in this way,
$\xi(r)$ is equivalent to the angle-averaged version of
\begin{equation}
\xi(\vec{r}) = \left<\delta(\vec{x})\,\delta(\vec{x}+\vec{r}) \right>,
\end{equation} 
and it also corresponds to the Fourier transform of the power
spectrum.  Note, however, that this estimate for $\xi(r)$ does not
require a shot-noise correction.

To accelerate the pair count, especially for large distances, we use a
tree-based neighbour finding that detects nodes that fall fully within
one of the logarithmic shells set-up for our $\xi(r)$ measurement and
then counts the particles in one go without having to refine the tree
any further. For the large particle numbers we have for some of our
samples (for example, for measuring the total matter autocorrelation
function in TNG300-1 this is in excess of 30 billion), it would be
overly expensive and unnecessary to determine neighbour counts for
each point. Instead, a random subset is sufficient to obtain a
measurement of $\xi(r)$ that is negligibly affected by subsampling
noise. We usually use a limit of $N_{\rm max} =10^5$ measurements of
neighbour counts to estimate $\xi(r)$, i.e. if the particle number in
the set is smaller than $N_{\rm max}$, all points are considered and
the pair counts are thus complete, otherwise we randomly down-sample
the selection of points by a factor $N_{\rm max} / N$.  Our results
are not sensitive to the choice of $N_{\rm max}$ if chosen
sufficiently high as we do here.  

We usually refrain from estimating sample variance errors for our
simulated correlation functions as these errors are typically small
and subdominant compared to systematic effects from finite resolution
and physics modelling. We note that measuring the correlation function
in real-space in this way, as opposed to trying to Fourier-transform a
measurement of the power spectrum, circumvents the thorny issue of
shot-noise corrections and is thus our preferred approach.

To compare to observational measurements of the correlation function
we usually employ the projected correlation function
\begin{equation}
w_p(r_p) = 2 \int_{0}^{\infty} \xi\left(\sqrt{r_p^2 + \pi^2}\,\right)
{\rm d}\pi 
,
\label{eqnwp}
\end{equation}
which integrates along the line of sight to remove effects from
redshift-space distortions \citep{Davis1983} in observational
determinations of galaxy clustering, where the
measured $\tilde \xi(r_p, \pi)$ is a function of both the transverse
distance $r_p$ and the line-of-sight separation $\pi$
\citep[e.g.][]{Fisher1994}, whereas we can directly measure the spherically
symmetric real-space correlation function $\xi(r)$ appearing in
equation~(\ref{eqnwp}).  We carry out the integration numerically,
extending it to $\pi_{\rm max} \sim 80\,{\rm Mpc}$. We note that
observational studies, sometimes need to restrict the integration to
smaller distances, especially at high redshifts, which can bias the
result low, for example by about 10\% when
$\pi_{\rm max} \sim 20\,{\rm Mpc}$ \citep{delaTorre2011}.

For comparing to the
linear theory autocorrelation function, we compute it from the
dimensionless power spectrum  through
\begin{equation}
\xi_{\rm lin}(r) = \int_0^\infty \Delta_{\rm lin}^2(k)
\frac{\sin(kr)}{kr}\,\frac{{\rm d}k}{k} ,
\label{eqnlinx}
\end{equation}
where $\Delta_{\rm lin}^2(k)$ is the linear theory input power
spectrum extrapolated to the redshift under consideration with the
linear growth factor.

\begin{figure*}
  \centering
\includegraphics[width=0.5\textwidth]{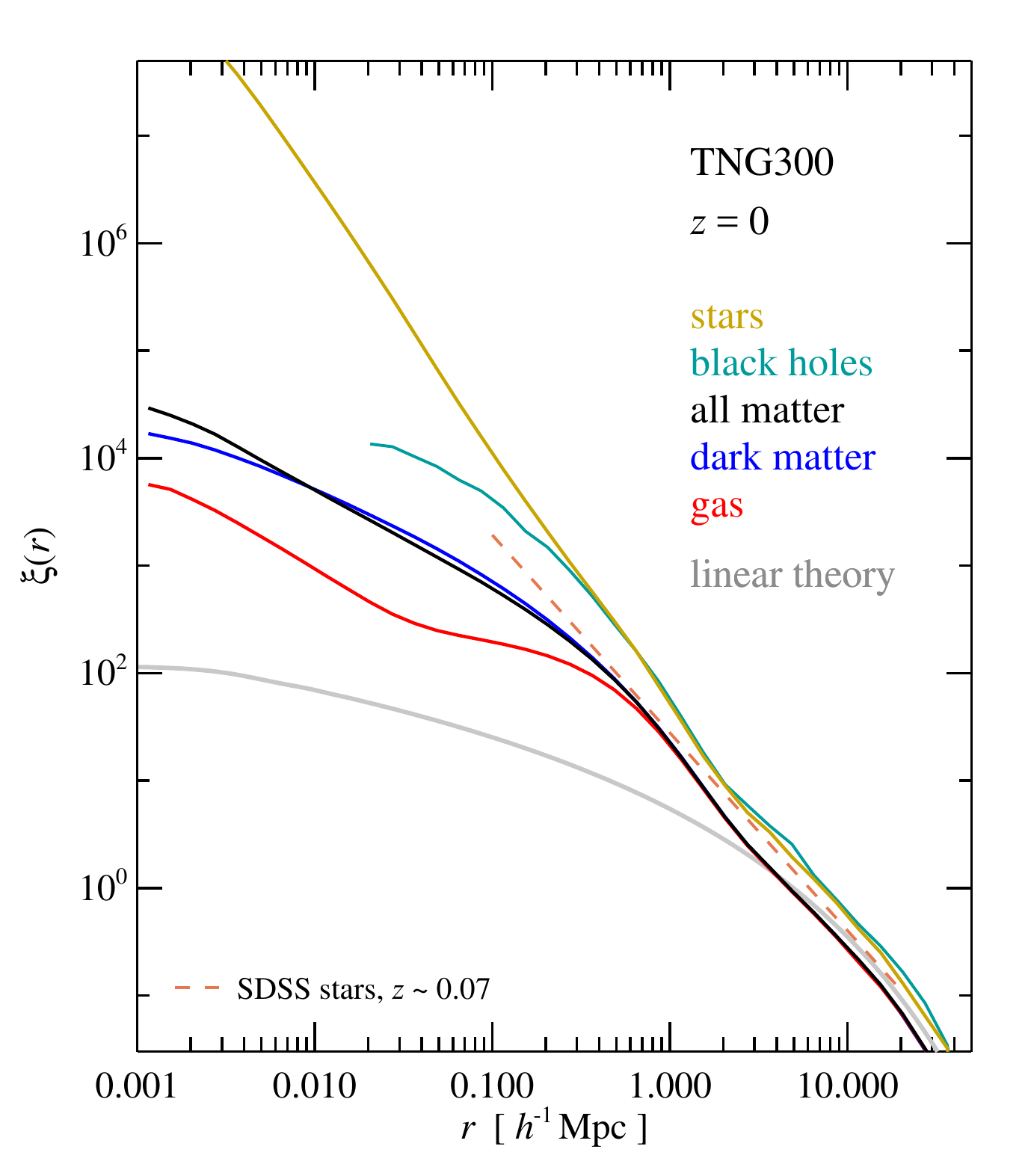}%
\includegraphics[width=0.5\textwidth]{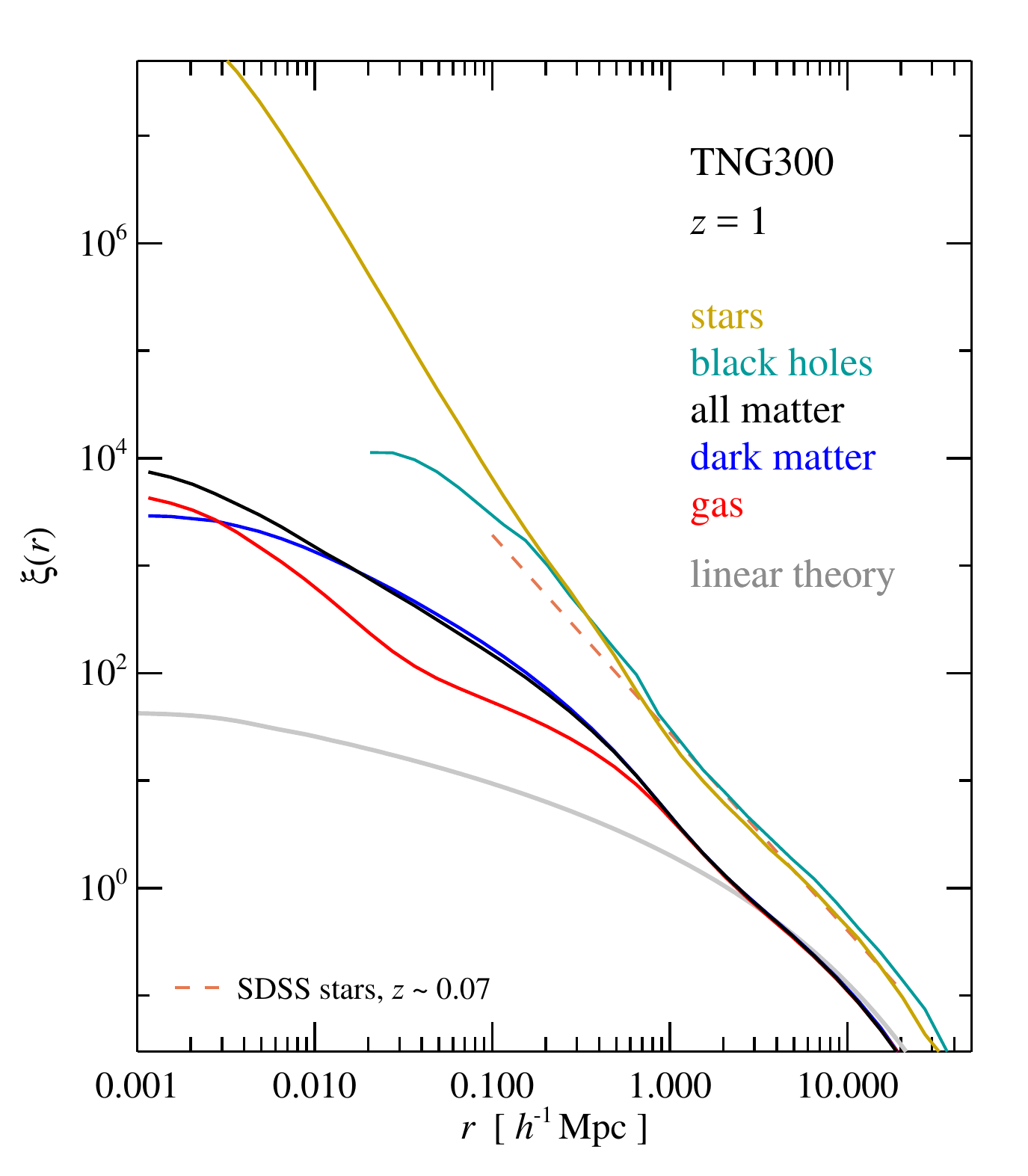}\\
\includegraphics[width=0.5\textwidth]{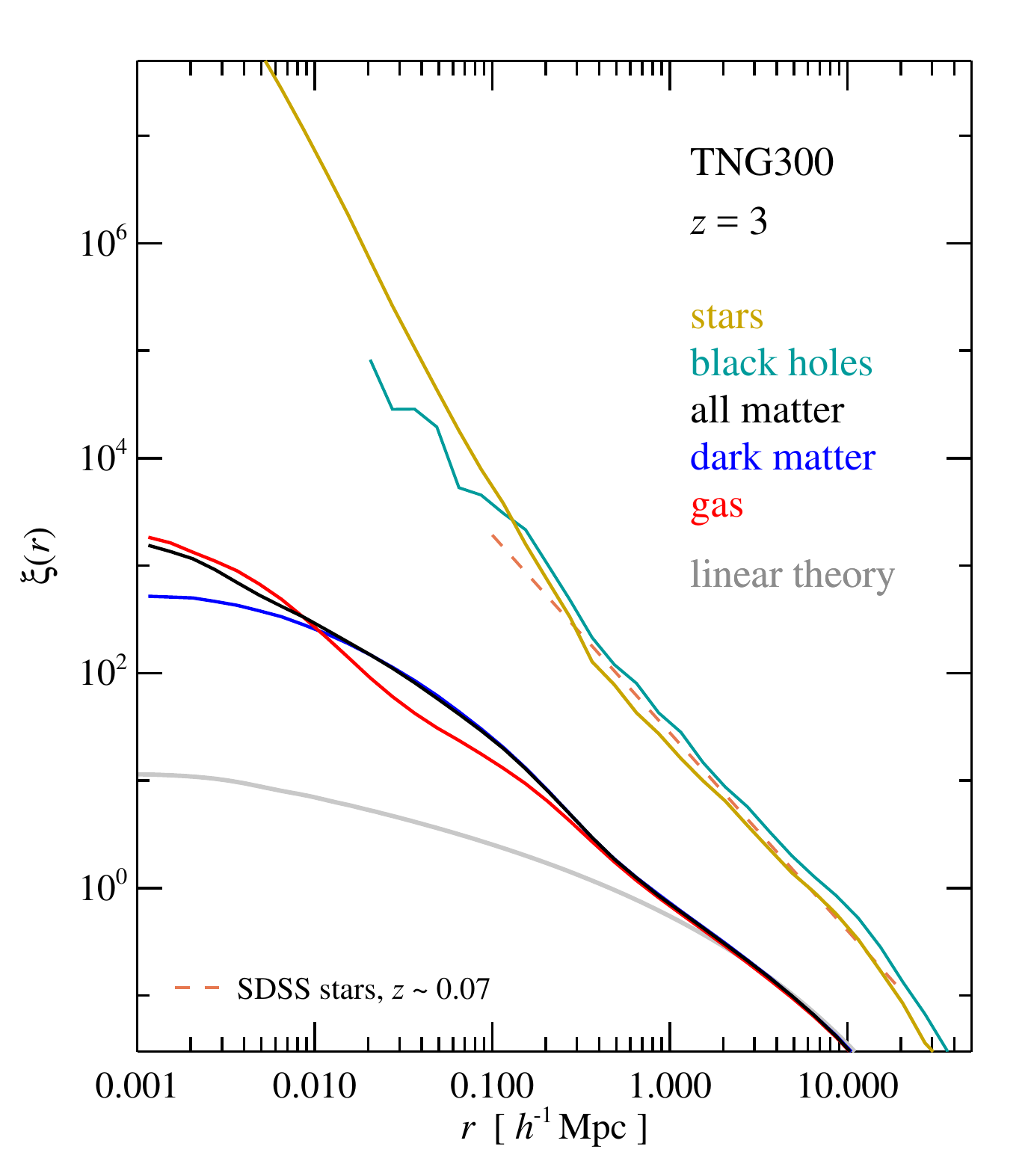}%
\includegraphics[width=0.5\textwidth]{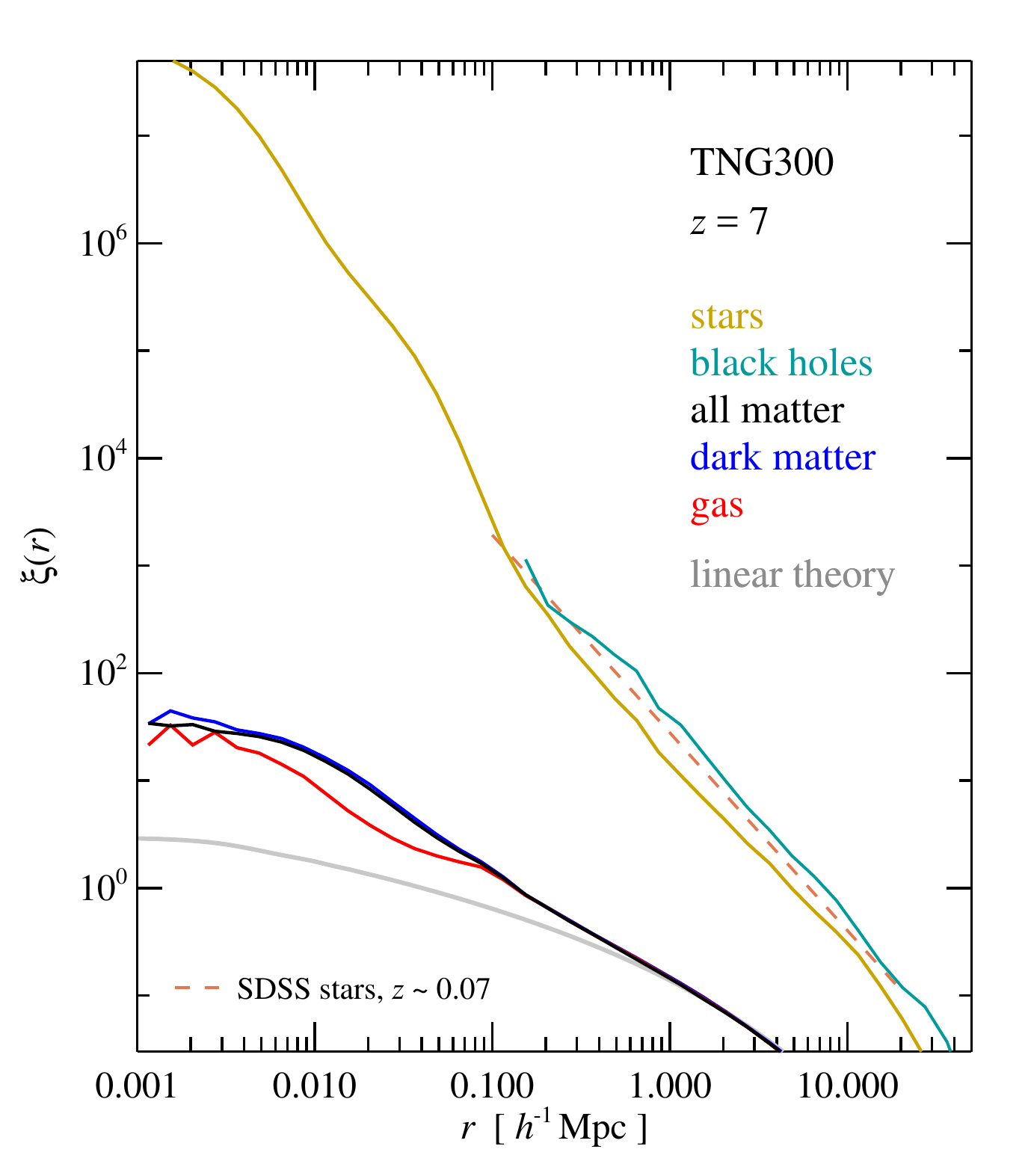}\\
\caption{The matter autocorrelation function for different mass
  components in our high-resolution TNG300 run at redshifts $z=0$,
  $z=1$, $z=3$ and $z=7$. We show results for stellar matter, 
  gas, dark matter, black holes, and all the matter, as labelled. The
  linear theory correlation function is shown in grey for
  comparison. The dashed line gives the autocorrelation function for
  all the stellar mass estimated by \citet{Li2009} for the low
  redshift Universe using nearly half a million galaxies from the
  Sloan Digital Sky Survey. This power law,
  $\xi_{\star}(r) = [r / (6.1\, h^{-1}{\rm Mpc})]^{-1.84}$, is
  reproduced in all the plots as a reference point.  }
  \label{fig:matter2cf}
\end{figure*}

\begin{figure*}
  \centering
\hspace*{0.8cm}\resizebox{!}{10.5cm}{\includegraphics{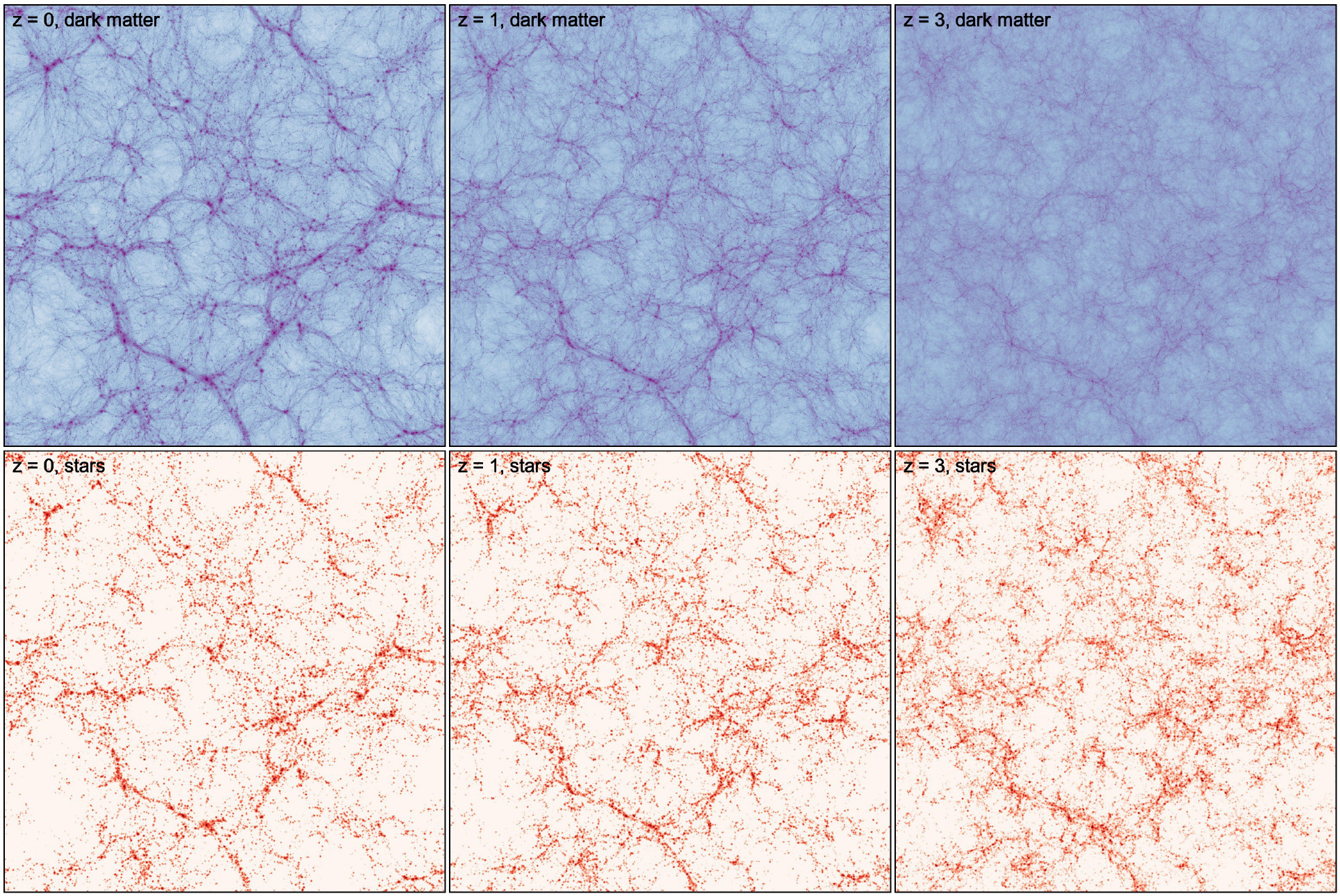}}%
\resizebox{!}{10.5cm}{\includegraphics{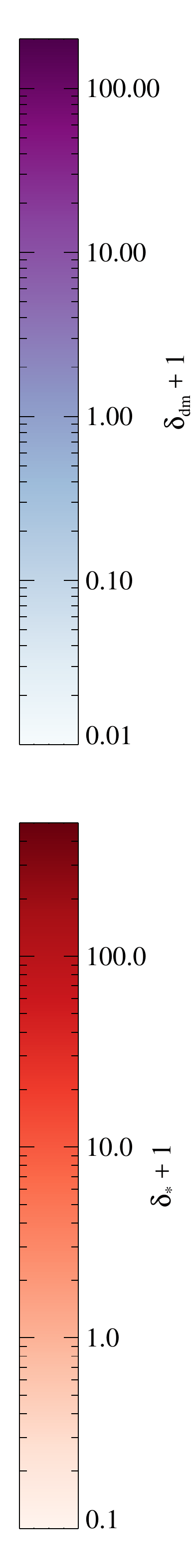}}\\
\caption{Projected dark matter and stellar density fields in TNG300,
  at redshifts $z=0$, $z=1$, and $z=3$. The slices are
  $205\,h^{-1}{\rm Mpc}$ wide (full width of the box) and
  $25\,h^{-1}{\rm Mpc}$ thick, with the density fields being
  normalised to the mean density in each panel. The density field of
  the stars has been smoothed with a Gaussian filter of width
  $160\,h^{-1}{\rm kpc}$ to make it more volume filling and hence
  better visible. While the density contrast in the dark matter
  distribution progressively increases with time, the clustering of
  the stellar matter is already strong at high redshift and evolves
  little with time.}
  \label{fig:projections}
\end{figure*}

\subsection{Bias measurement}

We also determine the clustering bias of different samples of galaxies or haloes
with respect to the total matter, both in real space and in Fourier
space. For example, when working in real space, we define
the ratio
\begin{equation}
b(r) = \left[\frac{\xi_{\rm gal}(r)}{\xi(r)}\right]^{1/2} 
\end{equation}
as the bias of a galaxy sample with measured correlation function
$\xi_{\rm gal}(r)$ relative to the total mass. Here $\xi(r)$ is the
(non-linearly) evolved correlation function for the total matter as
measured from the simulation. Sometimes the linear theory correlation
function is used instead for defining the bias, but this is expected
to amplify the scale-dependence of the bias which then also needs to
account even for mildly non-linear evolution of the clustering of
matter.

Similarly, when working in $k$-space, we define the bias 
as the ratio of the power spectra,
\begin{equation}
b(k) = \left[\frac{P_{\rm gal}(k)}{P(k)}\right]^{1/2}.
\end{equation}
On the largest scales represented in the box, we expect that the bias
factors $b(r)$ and $b(k)$ become equal and constant with scale,
something that we call the linear bias. To measure the linear bias, we
compute the average bias for the largest modes represented in the
simulation box, assuming that scale independence has been reached
there.  Where exactly scale-dependent effects set in is one of the
interesting questions that simulation models like TNG should help to
answer.

When measuring halo bias, we define the positions of haloes through the
locations of their potential minima, and their masses through the
spherical overdensity (SO) approach with a density contrast of 200
relative to the critical density. For comparison with literature results
for the linear bias on the largest scales, we adopt the often used
parameterisation of the halo mass in terms of peak height,
$\nu = \delta_c / \sigma(M)$, where $\delta_c = 1.686$ is the linearly
extrapolated overdensity for top-hat collapse, and $\sigma^2(M)$ is
the variance of the linearly extrapolated density field when filtered
with a top-hat filter containing the mass $M$, i.e.
\begin{equation}
\sigma^2(M) = \frac{1}{(2\pi)^3} \int P(k) |W_R(k)|^2 4 \pi k^2 {\rm d} k ,
\end{equation}
where $W_R(k)$ is the Fourier transform of the top-hat window of
radius $R$. This filter-scale is set such that a sphere of radius
$R$ contains the mass $M$ of the halo at mean background density
$\overline{\rho}$, i.e.~$M = (4\pi/3) \overline{\rho} R^3$.

The simulated galaxy samples we study are based on an identification
of locally overdense, gravitationally bound structures in the TNG
simulations with the {\small SUBFIND} algorithm
\citep{Springel2001}. We require objects to have at least a dark
matter mass fraction of 10\% in order to filter out a small population
of bound baryonic lumps that appear to be produced by disk
fragmentation. We do not distinguish between central and satellite
galaxies in this work. The stellar masses we assign to the simulated
galaxies and which are used in various cuts to select subsamples are
based on the measured stellar mass within twice the stellar half mass
radius of each subhalo, as described in more detail in the
documentation of our public data release of Illustris
\citep{Nelson2015}. We note that our clustering results are quite
insensitive to the adopted definition of galaxy stellar masses, and
hence are also hardly affected by lack of full convergence of the stellar
masses for low mass galaxies in TNG300.

\section{The clustering of matter}   \label{sec_matter}

In Figure~\ref{fig:matter2cf}, we show the two-point correlation
function of different matter components in the TNG300 simulation at
redshifts $z=0$, $1$, $3$ and $7$. We include results for the dark
matter, the gas distribution, the star particles (i.e. the stellar
mass), the black hole mass, and the total matter distribution. For
comparison, the linear theory two-point correlation function is given
at the corresponding redshifts as well.

Clear effects of non-linear evolution of the matter correlation
function are visible for $ r < 1\, h^{-1} {\rm Mpc}$ at $z=7$, and
they propagate with time to ever larger scales as the matter
correlation function develops a characteristic `shoulder' on small
scales. As far as the dark matter goes, this can be explained in terms
of the halo model \citep[see][for a review]{Cooray2002} where the
clustering signal on small scales, below $\sim 2\,h^{-1}{\rm Mpc}$, is
dominated by particle pairs in the same halo (`one-halo term'), and
the larger scale correlations come from pairs in different haloes
(`two-halo term'). {Note that at $z=0$ the non-linear $\xi(r)$ of
  the total matter distribution falls slightly {\it below} the linear theory
  $\xi_{\rm lin}(r)$ at quasi-linear scales around $5\,h^{-1}{\rm Mpc}$. This
  happens despite the fact that the non-linear power spectrum is
  always larger than or equal to the linear power spectrum, and can occur for
  a limited range of $r$ due to the oscillatory factor $\sin(kr)/kr$ in
  equation~(\ref{eqnlinx}). This effect of non-linear evolution has already
  been seen in early N-body simulations \citep[e.g.][]{Ma1999} and
  can be interpreted in physical terms as a reflection of the
  depletion of matter on quasi-linear scales due to gravitational
  infall onto halos. }

It is interesting that the baryonic gas starts to differ
from the dark matter in the one-halo regime already early on. At redshift
$z=3$, the gas is actually more clustered than the dark matter on the
smallest scales, while its clustering signal is slightly suppressed on
intermediate scales. This changes qualitatively at low redshift, where
the gas becomes less clustered than the dark matter also on small
scales, and the overall suppression relative to dark matter becomes
substantially larger. The strong small-scale clustering of the gas at
$z=3$, which dominates the matter power spectrum in this regime,
reflects the intense cooling and star formation rates at this epoch,
while the clustering deficit at late times is caused by the growing
population of quenched, gas-poor galaxies that have depleted their gas
reservoirs through star formation and expelled some of the baryons
from their host haloes by feedback effects. Interestingly, this is at
least qualitatively consistent with observational evidence for
galaxies being baryon-dominated in their inner regions at the peak of
galaxy formation activity \citep{Genzel2017}.

Another striking result is the very strong clustering of the stellar
mass, which at low redshift is quite close to a power-law correlation
function over a very large dynamic range. This clustering is nearly
invariant in time, and for scales $r \ge 1\,h^{-1}{\rm Mpc}$ agrees
very well with the power-law auto-correlation function of the stellar
mass inferred by \citet{Li2009} for the data release 7 of the Sloan
Digital Sky Survey (SDSS) at a redshift of $z\sim 0.1$, which is
reproduced as a dashed line in all the panels of
Fig.~\ref{fig:matter2cf}.  Although we find a clear steepening of the
stellar mass auto-correlation towards smaller scales, this can be
viewed as a first indication that the clustering of our simulated
galaxies is in reasonably good agreement with observations. Also, it
readily indicates that the bias of the stellar mass relative to the
dark matter is large at high redshift, and then declines with
time. This is also illustrated by the projected dark matter and
stellar density fields shown in Figure~\ref{fig:projections} for the
TNG300 simulation. The time evolution from $z=3$ to $z=0$ shows the
gradual emergence of an ever more prominent cosmic web out of an
initially nearly uniform dark matter distribution. In stark contrast,
the stellar mass density field is highly structured already early on
and evolves comparatively little with time.

\begin{figure}
  \centering
  \includegraphics[width=0.47\textwidth]{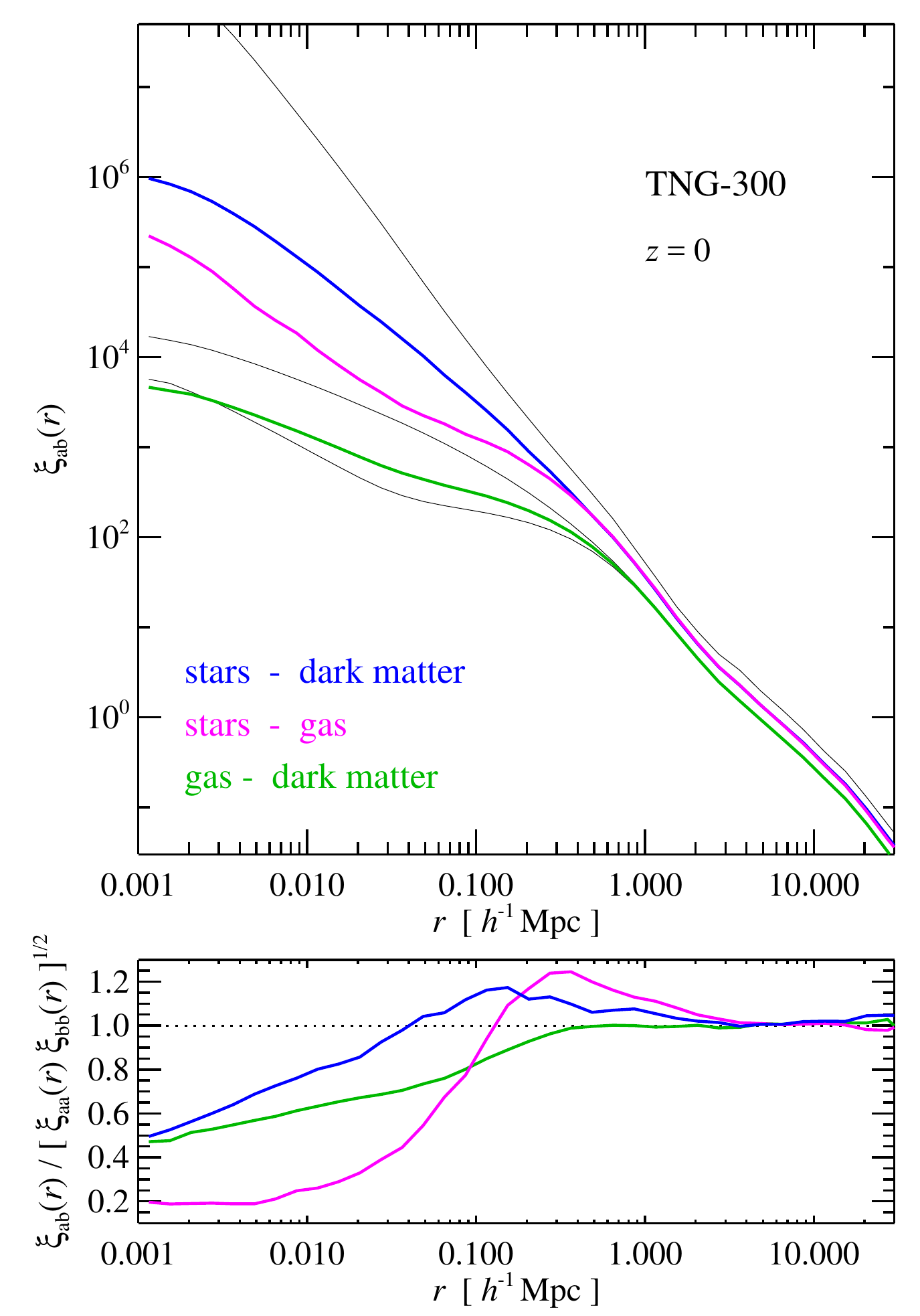}
  \caption{Matter cross-correlation functions $\xi_{ab}(r)$ in real space
between different mass components, where $a$ and $b$ stand for
stellar mass, dark matter or gas, respectively. The autocorrelation functions of stars, dark matter and gas (from top to bottom) are indicated with thin grey lines, for
reference. The bottom panel expresses the three cross-correlation functions
in units of the geometric mean of the auto-correlation
functions of the two involved matter fields. This pseudo-correlation
coefficient approaches unity only on large scales, showing that only
there a simple linear bias suffices to describe the relation between the two fields.
\label{fig:matter2cf_cross}}
\end{figure}

\begin{figure*}
  \centering
  \includegraphics[width=0.99\textwidth]{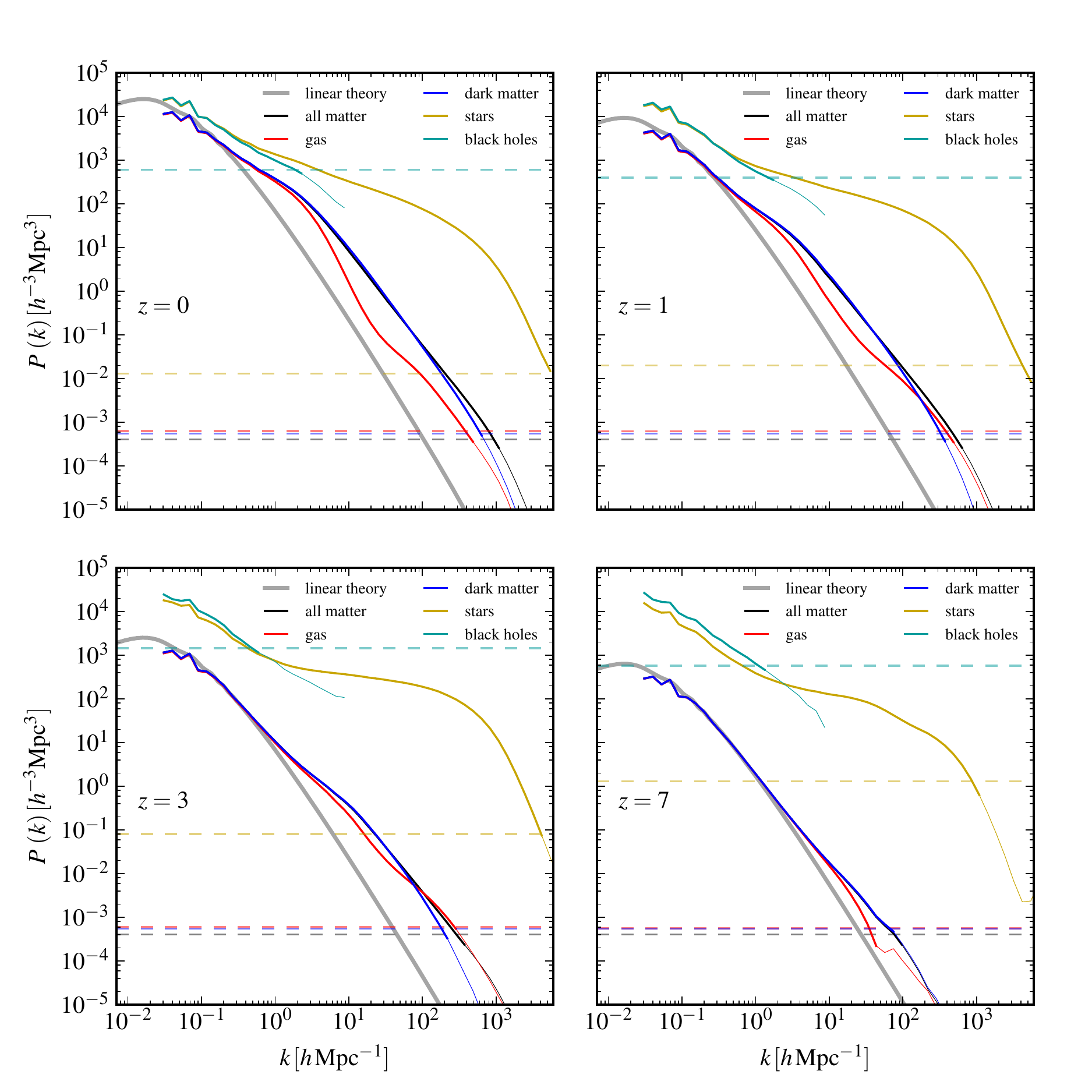}
  \caption{Matter power spectra for different mass components of the
    TNG300 simulation at redshifts $z=0$, $1$, $3$ and $7$, as
    labelled. The horizontal dashed lines in each panel give the
    formal shot-noise contribution of the corresponding mass
    component. The shot noise has been subtracted in all cases, and we
    continue to plot the obtained estimate for the underlying power
    spectrum below the shot noise limit, albeit with a thinner line
    style to indicate the uncertainty of this correction due to the
    fact that some of the tracers do not trace the underlying field in
    a perfectly Poissonian fashion. The grey lines show the linear
    theory power spectra at the corresponding redshifts.}
  \label{fig:simpower}
\end{figure*}

Figure~\ref{fig:matter2cf} further shows that the clustering of the
black hole mass in the simulations follows that of the stellar mass
closely at $z=0$, except on small scales, where the black hole
two-point correlation function starts to fall short at
$\sim 200\,h^{-1} {\rm kpc}$ and then suddenly drops to extremely low
values for scales below about $20\,h^{-1} {\rm kpc}$. This can be
understood from the rapid merging of black hole pairs in our
simulation model once they occupy the same halo. At the resolution of
our simulations, the sinking of black holes to the potential minima of
haloes due to dynamical friction cannot be followed accurately, hence
we reposition black holes to the potential minimum of their host halo
once they are close to the halo centre.  This effectively assumes that
dynamical friction is very efficient in bringing the black holes
together, and that black hole binaries are formed quickly and then
merge on a short time-scale. Towards higher redshifts, the
corresponding influence on the clustering of black holes is expected
to occur on slightly smaller scales due to the smaller sizes of haloes
there, consistent with our results. We note that we use the `internal'
black hole mass \citep{Springel2005BH} associated with the sink
particles for computing the clustering signal, not the sink's inertial
masses. These two can differ at high-redshift, where the seed black
hole masses are smaller than our nominal baryonic resolution.

The clustering signal of the black holes is dominated by the most
massive black holes, which have already grown significantly by gas
accretion, greatly helping to limit the dependence of our black hole
clustering results on the seeding prescription.  Interestingly, the
black hole mass exhibits a mild positive bias with respect to the
stellar mass towards high redshift. This can be interpreted as a
signature of top-down growth of black holes, where they first grow
preferentially in more massive haloes than the stars, and hence end up
being more strongly biased with respect to the matter. This difference
tends to vanish towards the present epoch, at which point a largely
universal ratio between stellar mass and black hole mass in galaxies
is established.

\begin{figure}
  \centering
\includegraphics[width=0.50\textwidth]{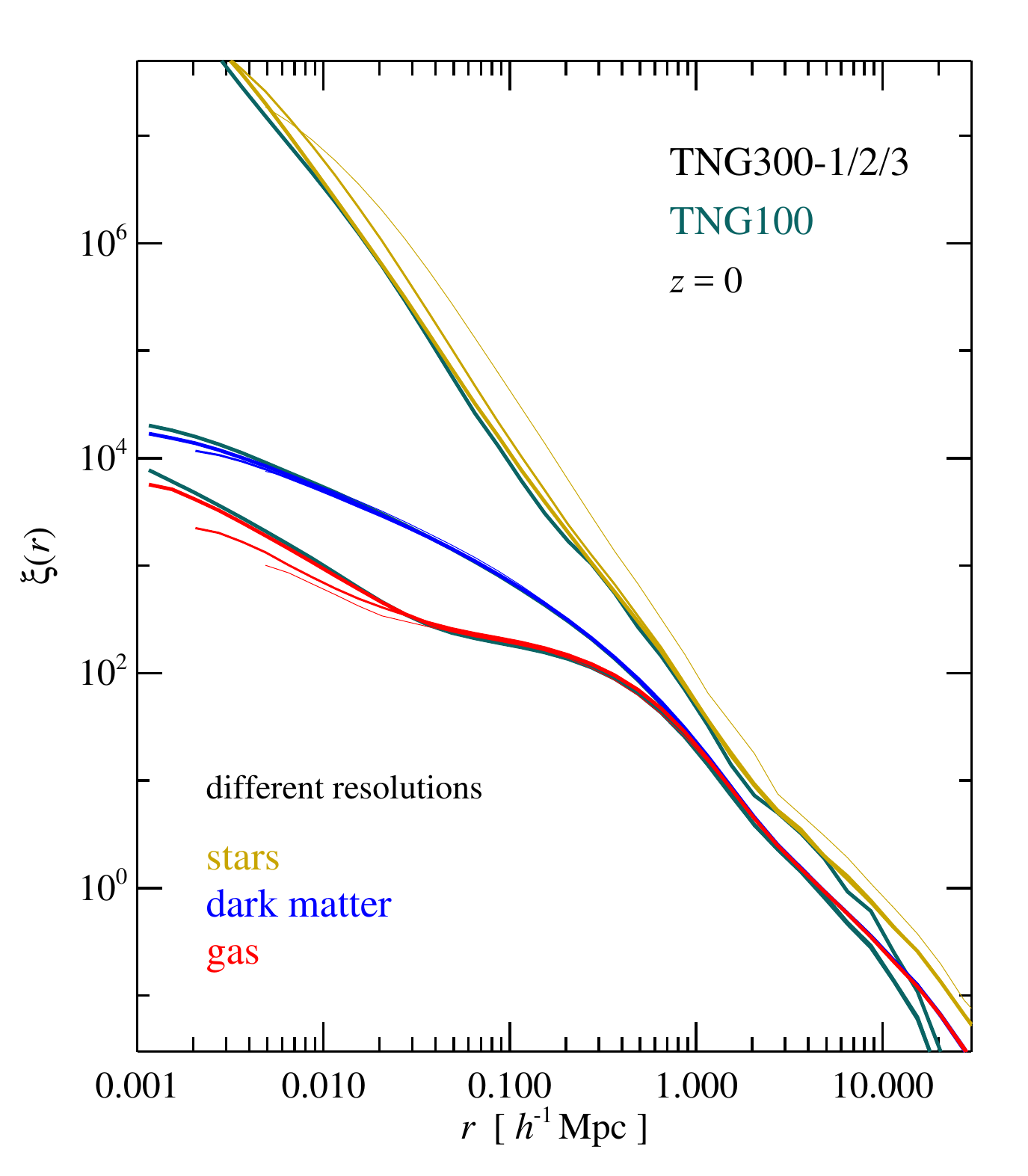}\\%
\includegraphics[width=0.50\textwidth]{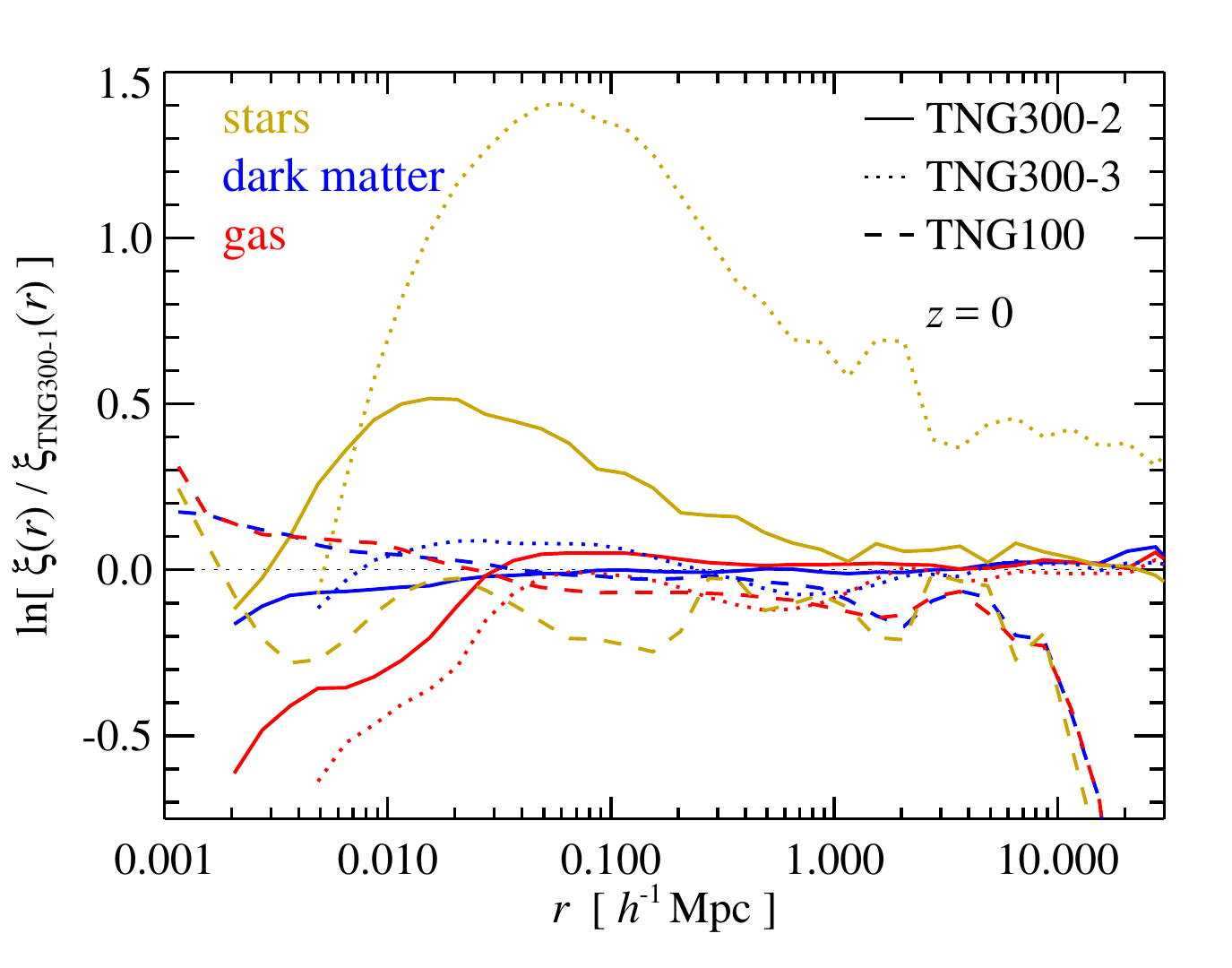}\\
\caption{Convergence of the auto-correlation functions in real space
for stellar matter, dark matter and gas. We show measurements for the \mbox{high-,} intermediate- and
low-resolution runs of TNG300, and also compare to the highest resolution run of TNG100, which
has about 8 times better mass resolution than TNG300-1.
The lines extend on small scales to each run's gravitational
softening length.
In the upper panel, the
thickest linestyle corresponds to the highest resolution TNG300-1
model, with the lower resolution versions TNG300-2/3 shown with progressively thinner line styles.
The TNG100 run is displayed with
turquoise thick lines. In the lower panel, we show the relative differences of the simulations relative to TNG300-1, as labelled. The convergence between TNG100 and
the highest resolution TNG300 run is rather good, even for the
auto-correlation function of the stellar mass. For distances
beyond $\sim 5\,h^{-1}{\rm Mpc}$, TNG100 shows a significant (and expected) deficit of
clustering strength due to its limited box size.
\label{fig:matter2cf_convergence}}
\end{figure}

In Figure~\ref{fig:matter2cf_cross}, we consider the cross correlation
functions between dark matter, gas and stars/black-holes\footnote{In
  this plot, we add the black hole mass to the stellar mass, for
  simplicity.} at $z=0$. The measured
correlation functions fulfil the relationship
\begin{eqnarray}
\rho^2 \xi(r)  & = & \rho_{\rm dm}^2 \xi_{\rm dm}(r) + 
 2 \rho_{\rm dm} \rho_{\rm gas} \xi_{\rm dm, gas}(r) \nonumber\\
& +  & \rho_{\rm gas}^2 \xi_{\rm gas}(r) + 2 \rho_{\rm dm}
       \rho_{\star} \xi_{\rm dm, \star}(r) \nonumber \\
& + &   \rho_{\star}^2 \xi_{\star}(r)  + 2 \rho_{\rm gas} \rho_{\star} \xi_{\rm gas,\star}(r),
\end{eqnarray}
by construction. 
The measurements demonstrate that all three
mass components generally trace each other well, particularly on large
scales. This can also be explicitly verified by considering
a generalised correlation coefficient, such as the ratio
\begin{equation}
  \kappa_{\rm dm, gas} (r) = \frac{  \xi_{\rm dm, gas}(r)}{\sqrt{  \xi_{\rm dm}(r) \xi_{\rm gas}(r)}}
\end{equation}
for dark matter and gas, and similarly for other pairs of matter
components, as shown in the lower panel of
Figure~\ref{fig:matter2cf_cross}. For a simple linear bias, we expect
$\kappa\sim 1$, which is indeed achieved for all pairs of matter
components on large scales. On smaller scales, the degree of
correlation between the different fields becomes however
weaker. Interestingly, at $z=0$, the stars correlate better with dark
matter than gas on halo scales and below, probably a
reflection of the stronger alignment of the centrally concentrated
distributions of stars with the dark matter cusps.

In Figure~\ref{fig:simpower}, we consider the power spectrum results
for the same set of redshifts as shown in In
Fig.~\ref{fig:matter2cf}. The qualitative behaviour of the different
mass components is consistent with the real-space clustering discussed
earlier. While the power spectra of the stars, black holes and the
dark matter show comparatively little evolution between $z=1$ and
$z=0$, the gas actually shows a decrease in power at small and
intermediate scales. This implies a non-monotonic evolution of the gas
clustering with time, which can be interpreted as a signature of
strong late-time feedback effects in the gas distribution. These results
should be very informative for attempts to model the non-linear matter
power spectra of stars and the gas phase analytically through extensions
of the halo model \citep{Fedeli2014a, Fedeli2014b}.

\begin{figure*}
  \centering
  \includegraphics[width=1.04\textwidth]{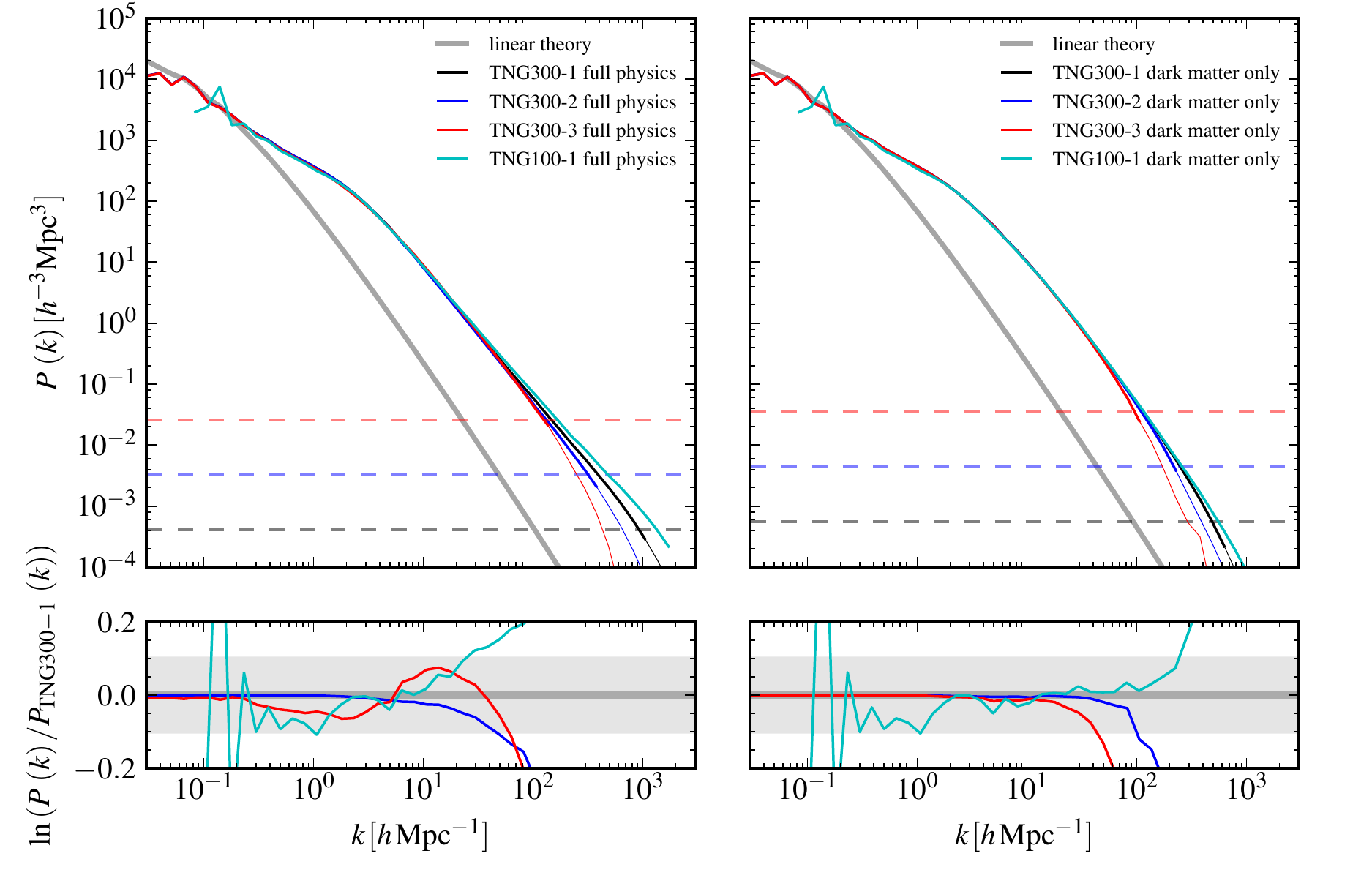}
  \caption{Resolution dependence of the total matter power spectrum in
    the full physics runs (left panel), and the dark matter only runs
    (right panel) at $z=0$. In both cases, we show results for the
    TNG300-1/2/3 simulations and the TNG100 level-1 simulation.
    The horizontal lines give the formal shot noise limits for
    variable mass Poissonian tracers of an underlying density
    field. This shot noise has been subtracted in all
    measurements. The grey lines indicate the linear theory power
    spectrum, for comparison. The bottom panels show the differences
    to the TNG300-1 run in terms of the natural logarithm of the power
    ratio, which allows to easily read off relative differences in
    percent.  The grey bands in the bottom panels denote differences
    of $10\%$ (light) and $1\%$ (dark), respectively.}
  \label{fig:simres}
\end{figure*}

We examine the resolution dependence of our two-point correlation
function estimates for the full physics simulation in
Figure~\ref{fig:matter2cf_convergence}, separately for the stellar,
dark matter and gaseous mass components.  The clustering signal of the
gas, dark matter, and stars is robustly reproduced even when varying
the mass resolution by a factor of more than $500$ between TNG100-1
and TNG300-3, except for an excess of the clustering of the stars in
the lowest resolution TNG300-3 simulation. In this calculation, star
formation in low mass haloes is poorly resolved and anaemic, so that
stars occupy preferentially more massive and rarer haloes that are
more strongly biased.  There are also some small differences between
the runs close to the spatial resolution limits, which are of the
expected magnitude. More importantly, the TNG100 model shows a
significant deficit of clustering at very large scales, already
setting in for $r>5\, h^{-1}{\rm Mpc}$. This is due to the limited box
size of this simulation, which clearly affects the clustering on
scales typically probed in galaxy surveys. The impact of the limited
box size can be estimated by computing $\xi(r)$ through equation
(\ref{eqnlinx}), but restricting the integration to $k\ge 2\pi/L$,
i.e.~modes represented in the box. This shows that the linear theory
two-point correlation function of the TNG100 box size is expected to
turn negative by $r\simeq 20\,h^{-1}{\rm Mpc}$, whereas the
TNG300 simulation is affected only by 10\% on this scale, and by much
less on smaller scales. The correlation function of TNG300 turns
negative at a scale of $r\simeq 50\,h^{-1}{\rm Mpc}$.

A corresponding analysis of the resolution dependence of the total
matter power spectrum is given in Figure~\ref{fig:simres}, with the
left panel focusing on our full physics simulations while the right
panel considers the corresponding dark matter only simulations.  On
large scales the power spectra agree very well, to better than 1\% for
scales down to a few times $0.1\,h\,{\rm Mpc}^{-1}$ for the full
physics simulations, and down to $\sim 10.0 \,h\,{\rm Mpc}^{-1}$ for
the dark matter only runs.

\begin{figure*}
  \centering
  \includegraphics[width=0.99\textwidth]{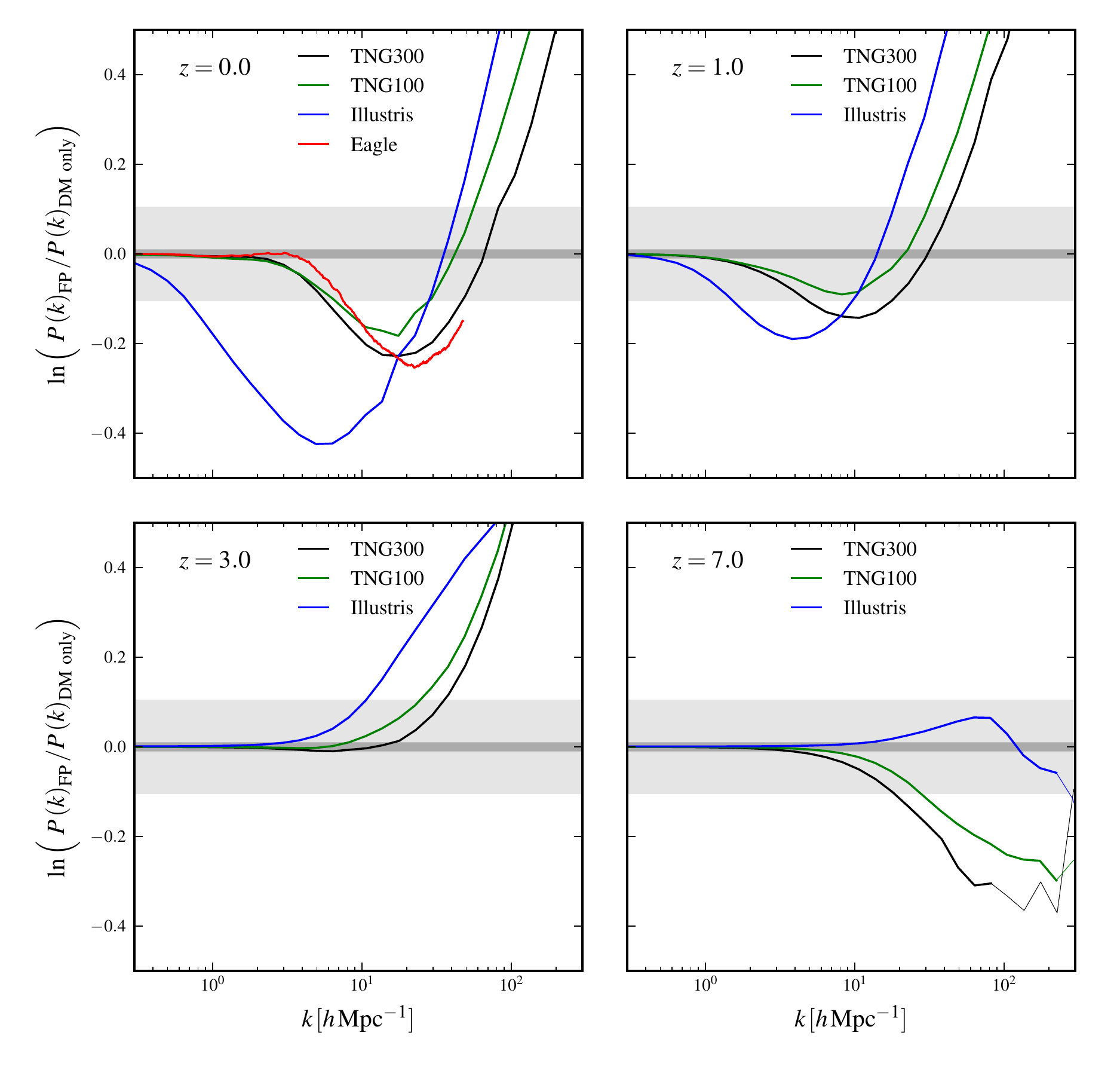}
  \caption{The ratio between the \textit{total matter} power spectrum
    of different full physics runs and the total power spectrum of
    their dark matter only companion runs at different redshifts, as
    labelled. We show results for TNG300, TNG100, and the Illustris
    simulation, and at $z=0$ also include a measurement for EAGLE by
    \citet{Hellwing2016}. We plot the natural logarithm of the ratio
    to allow easily reading off relative differences in percent.  The
    light and dark grey areas indicate relative differences of $10\%$
    and $1\%$, respectively.}
  \label{fig:FPvsDM}
\end{figure*}

Given the numerical robustness of the large-scale clustering results,
it is interesting to examine the overall impact of baryonic physics on
the clustering of matter, which is arguably one of the most
interesting effects that can be studied with hydrodynamic simulations,
as highlighted first by \citet{vanDaalen2011} and
\citet{Semboloni2011}. In Figure~\ref{fig:FPvsDM} we show the matter
power spectrum of TNG300 relative to the corresponding one of the
DM-only simulation, at redshifts $z=0$, $1$, $3$, and $7$.  For
comparison, we also include results for TNG100 and Illustris, as well
as for the EAGLE simulation at $z=0$ \citep{Hellwing2016}. At the
present epoch, the total change of the matter power spectrum is
described by a characteristic suppression of power by $\sim 20\%$ at
scales of $k\sim 10\,h\,{\rm Mpc}^{-1}$, and a strong and rapidly
rising enhancement of power setting in at scales around
$\sim 100\,h\,{\rm Mpc}^{-1}$. The effect we see in IllustrisTNG is
noticeably weaker than in Illustris, where the suppression extends to
considerably larger scales (the scale for which the power is
suppressed by more than 10\% is almost an order of magnitude larger),
and is stronger in amplitude, too. Interestingly, however, the effect
in TNG is qualitatively similar to the EAGLE simulation
\citep{Hellwing2016}, although it still is a bit stronger and extends
to slightly larger scales. This is despite the fact that we use
fundamentally different feedback prescriptions and numerical
techniques, suggesting that the size of the expected AGN feedback
impact is surprisingly robust to details of the modelling.

At higher redshifts, there are also some striking differences between
Illustris and IllustrisTNG. Apparently, the modified AGN model
and different wind parameterisation in these two simulations also
affects the timing when the suppression of power on intermediate
scales develops. In Illustris this emerges later than in the
IllustrisTNG model.

\begin{figure}
  \centering
  \includegraphics[width=0.47\textwidth]{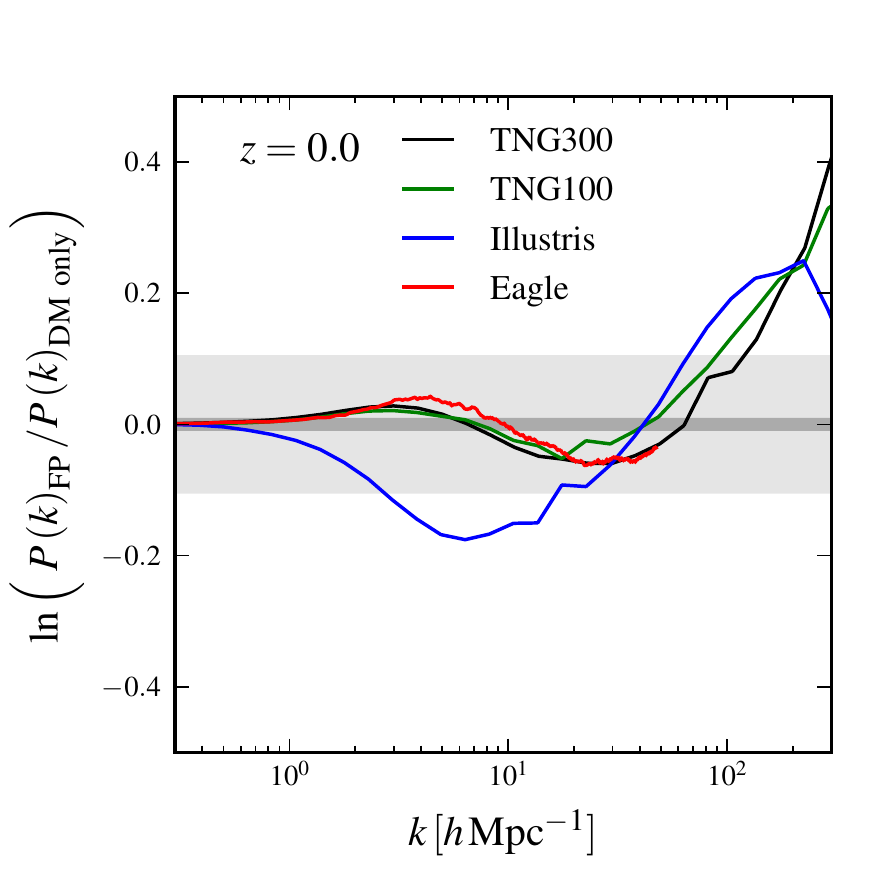}
  \caption{Impact of baryonic physics on the \textit{dark matter}
    power spectrum of different full physics runs at $z=0$. This is
    shown through ratios of the dark matter power spectrum to the
    power spectrum of the corresponding dark matter only
    simulation. We include results for TNG300, TNG100, and Illustris,
    as well as a measurement for the EAGLE simulation by
    \citet{Hellwing2016}. The light and dark grey areas denote
    variations of $10\%$ and $1\%$, respectively.}
  \label{fig:DMvsDM}
\end{figure}

The modification of the total matter power spectrum in the full
physics simulation is in part due to a redistribution of baryons by
non-gravitational physics, and in part due to a change of the dark
matter distribution as a result of the gravitational coupling to the
baryons. In Figure~\ref{fig:DMvsDM}, we look at the latter effect in
isolation, at $z=0$. The modification of the dark matter distribution
alone is sizeable but overall weaker than that of the total matter,
showing that the drastic change in the baryon distribution relative to
the dark matter brought about by galaxy formation physics is a primary
factor in determining the change of the total matter power
spectrum. Interestingly, the dark matter clustering not only shows a
damping on intermediate scales of $k\sim 30\,h\,{\rm Mpc}^{-1}$, but
also an enhancement of a few percent on $\sim 10$ times larger scales,
around $k\sim 3\,h\,{\rm Mpc}^{-1}$, where Illustris is still damped.
The latter effect is nearly twice as large in EAGLE than in
IllustrisTNG, but qualitatively the two simulations are relatively
similar, and exhibit a significant difference to the much stronger
effects in Illustris.

\begin{figure}
  \centering
  \includegraphics[width=0.47\textwidth]{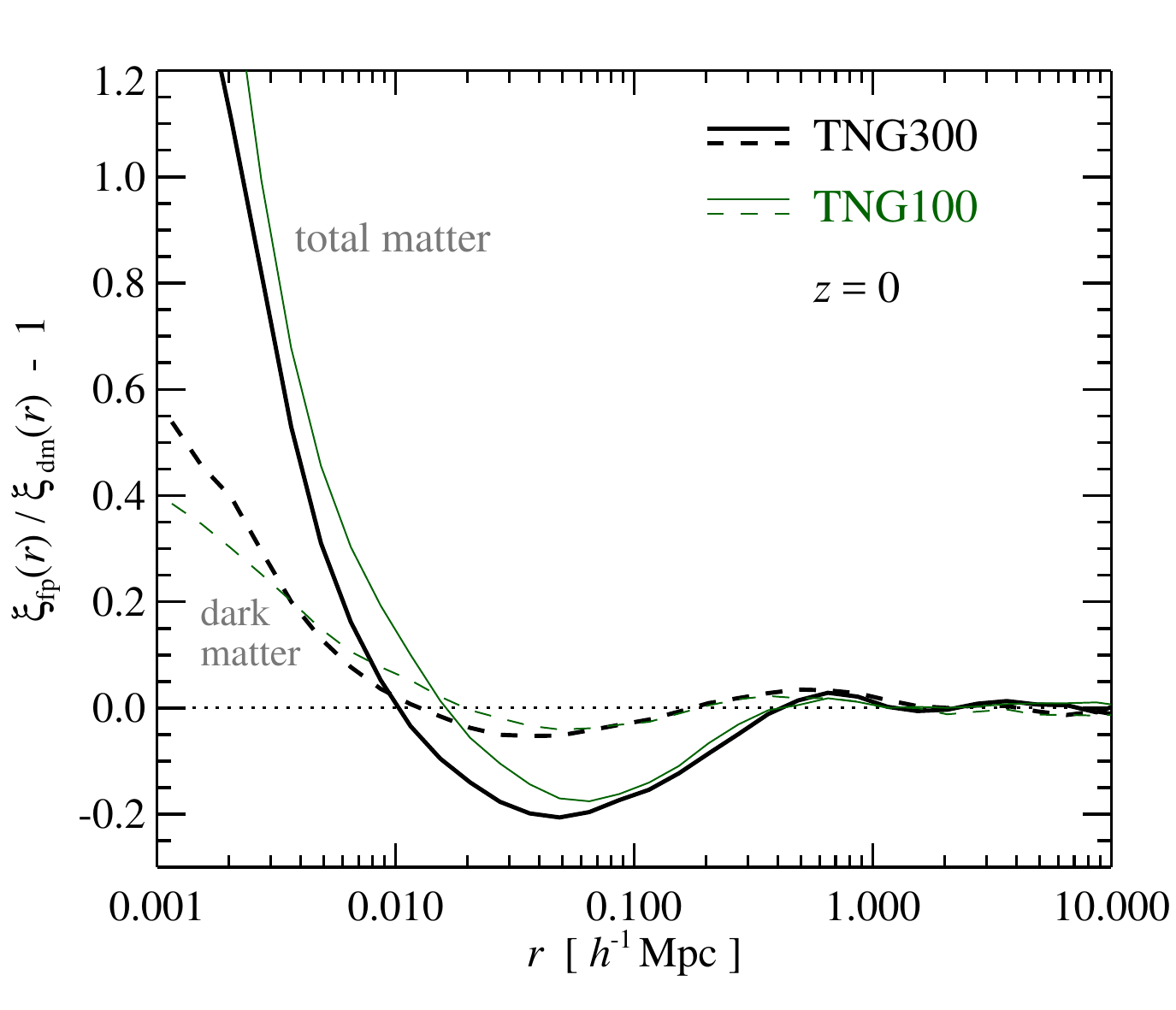}
  \caption{Impact of baryonic physics on the total and dark matter
    auto-correlation function in TNG300 and TNG100 at the present
    day. The measurements shown compare the full physics correlation
    function of all the mass (solid) and of just the dark matter
    (dashed) to the correlation function of the corresponding dark
    matter only simulation. }
  \label{fig:matter2cf_baryonicimpact}
\end{figure}

We can also consider the impact of baryonic effects on the two-point
correlation function \citep[see also][]{vanDaalen2014}, which is shown in
Figure~\ref{fig:matter2cf_baryonicimpact} for TNG300 and TNG100 at
$z=0$. The predictions of both simulations agree very well given their
substantial resolution and box size size differences. The solid lines
report the relative change of the total matter clustering with respect
to the clustering of the corresponding dark matter only
simulation. The full physics simulations show a suppression of the
clustering signal by about 20\% on scales of
$20-100 \,h^{-1}{\rm kpc}$, and a mild increase by about 5\% at around
$800\,h^{-1}{\rm kpc}$. Part of these changes are due to a
modification of the dark matter clustering itself, as shown by the
dashed lines, but the relative clustering difference of the baryons is
responsible for the bulk of the effect on small scales. At a distance
scale of $1\,h^{-1}{\rm kpc}$, the clustering of dark matter is
increased by approximately 40\% in the full physics calculations,
while the total clustering strength is already more than twice
as strong than in the corresponding dark matter only simulations.

\section{The clustering of galaxies}  \label{sec_galaxies}

IllustrisTNG predicts galaxies directly, in terms of gravitationally
bound groups of stars that are identified by the {\small SUBFIND}
algorithm \citep{Springel2001}. For each of these galaxies, we have
obtained measurements of basic properties such as stellar mass,
luminosity in different filter bands, morphology, size, or chemical
abundances. Studying these properties of the predicted galaxy
population lies traditionally at the heart of analysing hydrodynamical
simulations of galaxy formation. Much less attention has thus far been
given to analysing galaxy clustering in such simulations, in part due
to the box size limitations discussed above.

\begin{figure*}
  \centering
  \includegraphics[width=0.99\textwidth]{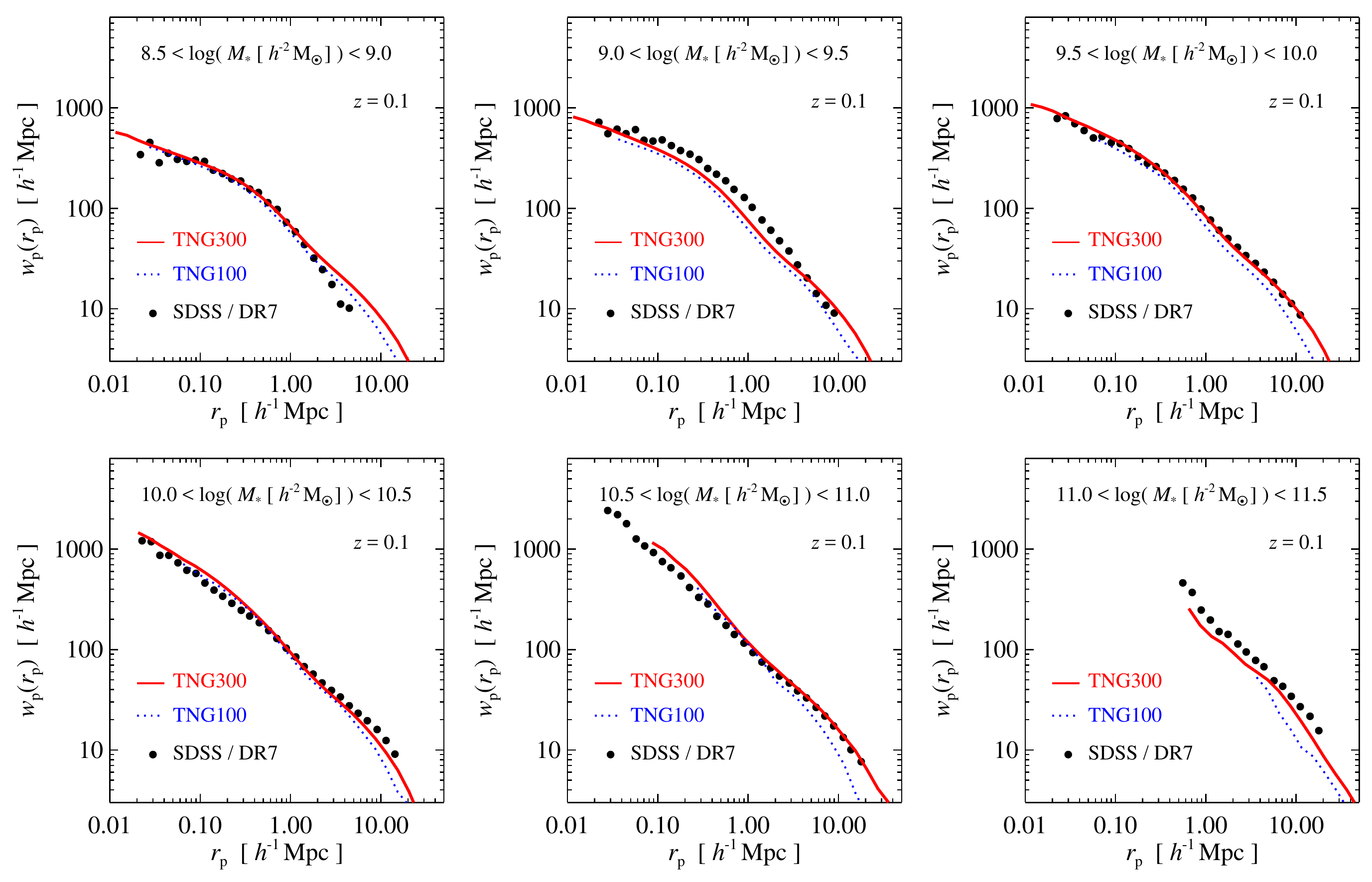}
  \caption{Comparison of the projected two-point galaxy correlation
    functions of TNG300 (solid) and TNG100 (dotted) at $z=0.1$ to the
    Sloan Digital Sky Survey, in six different stellar mass
    ranges. The data is taken from \citet{Guo2011} and
    \citet{Henriques2016}. Overall, the agreement is remarkably good,
    at about the same level of the currently best Munich
    semi-analytic galaxy formation model \citep{Henriques2016}. }
  \label{fig:2pcf}
\end{figure*}

This makes it all the more interesting to consider the clustering of
galaxies selected according to different criteria in our large volume
TNG300 simulation, and to compare it to observational constraints from
galaxy surveys.  In Figure~\ref{fig:2pcf} we compare the galaxies at
redshift $z\simeq 0.1$ to data from the Sloan Digital Sky Survey
(SDSS), as compiled by \citet{Guo2010, Guo2011}.  The SDSS provides by
far the most accurate characterisation of the low redshift galaxy
distribution \citep{Zehavi2011}, and thus imposes stringent
constraints on any galaxy formation model. We compare 6 bins of
stellar mass to SDSS, finding in general rather reassuring agreement
for the projected correlation functions, with a match of comparable
quality to that of current physically based semi-analytic models of
galaxy formation \citep{Henriques2016, Kang2014} or stochastic HOD
models \citep{Zu2017}. Also note that TNG300 and TNG100 agree rather
well, except for large scales, $r\sim 10\,h^{-1}{\rm Mpc}$, where
TNG100 lies noticeably lower (an effect that is however expected due
to the box size limitation of this simulation), and for the largest
stellar mass sample, where we have too few galaxies in TNG100 to
measure the correlation function for small separations. We are not
aware that a similar degree of agreement in the clustering data over a
comparably large dynamic range has ever been found for another
hydrodynamic simulation of galaxy formation \citep[see][for one of the
most succcessful other models]{Artale2016}.

\begin{figure*}
  \centering
  \includegraphics[width=0.99\textwidth]{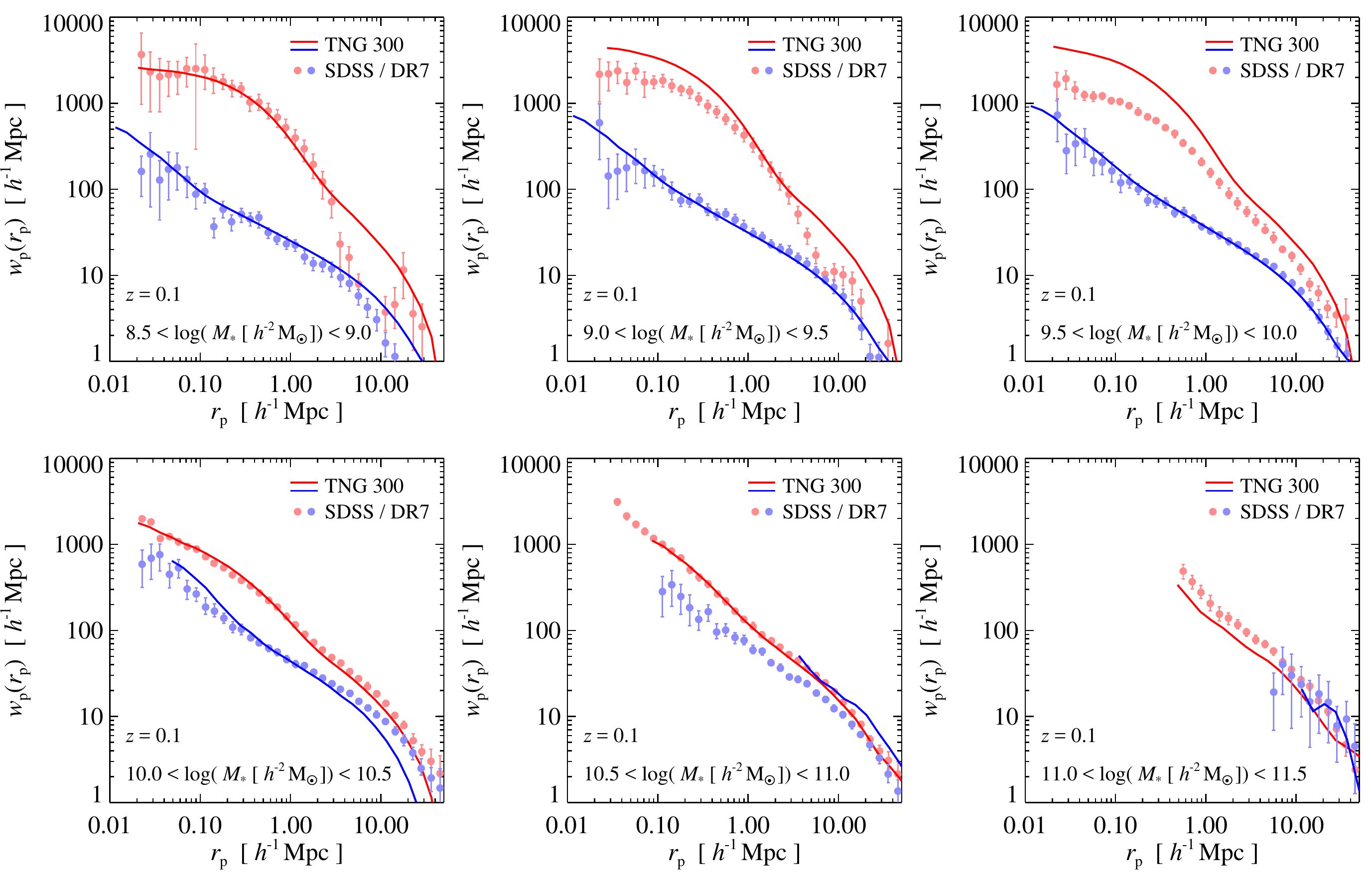}
  \caption{Projected galaxy correlation functions at $z=0.1$ split by
    $g-r$ color, for TNG300 and the Sloan Digital Sky Survey DR-7.  We
    show results for six different stellar mass ranges, as labelled,
    with data points taken from \citet{Guo2011} and
    \citet{Henriques2016}. The agreement for blue galaxies is
    generally very good. This is also the case for red galaxies,
    except in the stellar mass range
    $10^9-10^{10}\,h^{-2}{\rm M}_\odot$, where the simulation model
    shows a mild clustering excess. No dust corrections have been
    applied to the simulated galaxy colours, which could potentially
    alleviate this discrepancy.}
  \label{fig:2pcf-color}
\end{figure*}

We extend this comparison by splitting up the samples in terms of
galaxy colour in Figure~\ref{fig:2pcf-color}, using the cut
\begin{equation}
g - r = \log(M_\star / [h^{-1}{\rm M}_\odot]) \times 0.054 + 0.05  
\end{equation}
to distinguish between red and blue galaxies. Colors are assigned
using \citet{Bruzual2003} stellar populations synthesis models
assuming a Charbrier IMF. This lies at the bottom of the valley
separating the two populations and is similar to
\citet{Henriques2016}, but does not make any attempt to include dust
corrections. The match to the blue galaxies is excellent, essentially
for all stellar masses. The clustering of red galaxies appears
slightly overestimated for intermediate stellar masses. We note
however that the detailed size of this discrepancy depends on where
the colour split is taken \citep[see also the companion paper
by][]{Nelson2017TNG}, so we think this difference needs to be taken
with a grain of salt. In general, however, we consider the level of
agreement reassuring. It suggests that the quenching physics that
operates in our hydrodynamical simulations in a self-consistent manner
can broadly account for the observed clustering levels of red and blue
galaxies, and their detailed variations with stellar mass, which is a
non-trivial success.  At the same time, the small residual differences
can be used in the future to test extensions or modifications of the
physics model implemented in the simulations.

\begin{figure*}
  \centering
  \includegraphics[width=0.99\textwidth]{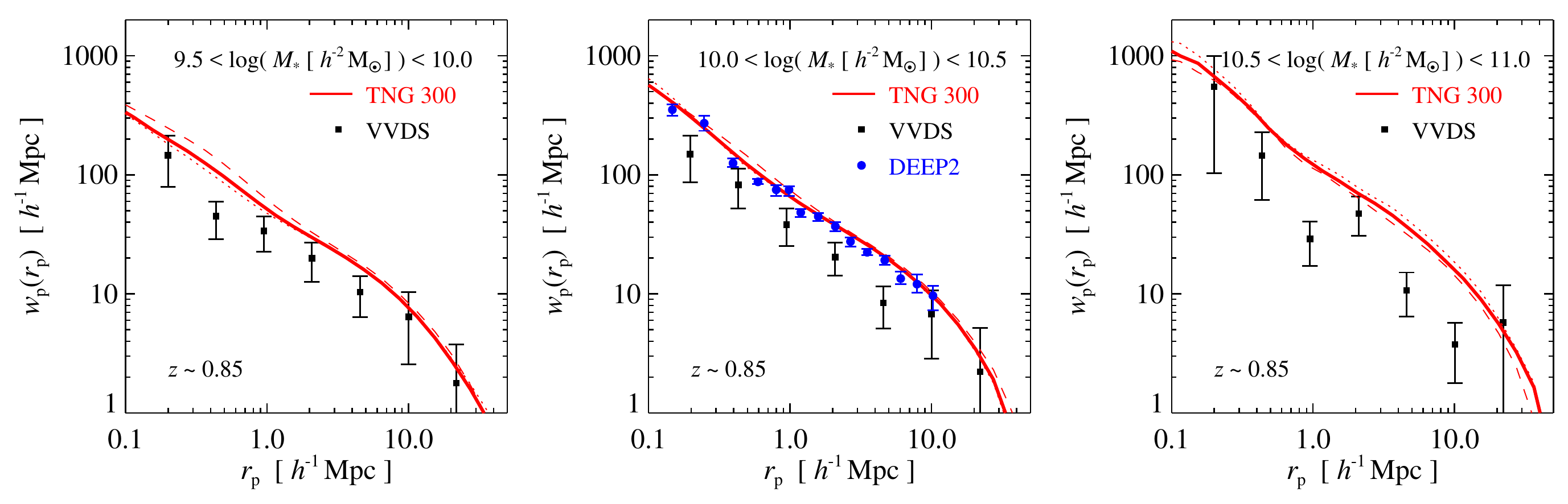}
  \caption{Projected galaxy correlation function of TNG300 in
    different stellar mass ranges at $z=0.85$ (solid thick lines),
    compared to data from the VIMOS VLT Deep Survey
    \citep[VVDS,][]{Meneux2008} and from the DEEP2 galaxy redshift
    survey \citep{Mostek2013}. The VVDS survey covers an extended
    redshift range, $0.5 < z < 1.2$, and we compare to the simulation
    results at the midpoint of this interval. To give an illustration
    of the very small variation of the simulation predictions over
    this time span, we also include TNG300 results for redshifts
    $z=0.5$ (dotted) and $z=1.2$ (dashed). The DEEP2 results are for a
    characteristic redshift $z\simeq 0.9$ and refer to an essentially
    complete sample of galaxies with
    $\log(M_\star / [h^{-2} {\rm M}_\odot]) > 10.16$.}
  \label{fig:2pcf-vvds}
\end{figure*}

\begin{figure*}
  \centering
  \resizebox{9cm}{!}{\includegraphics{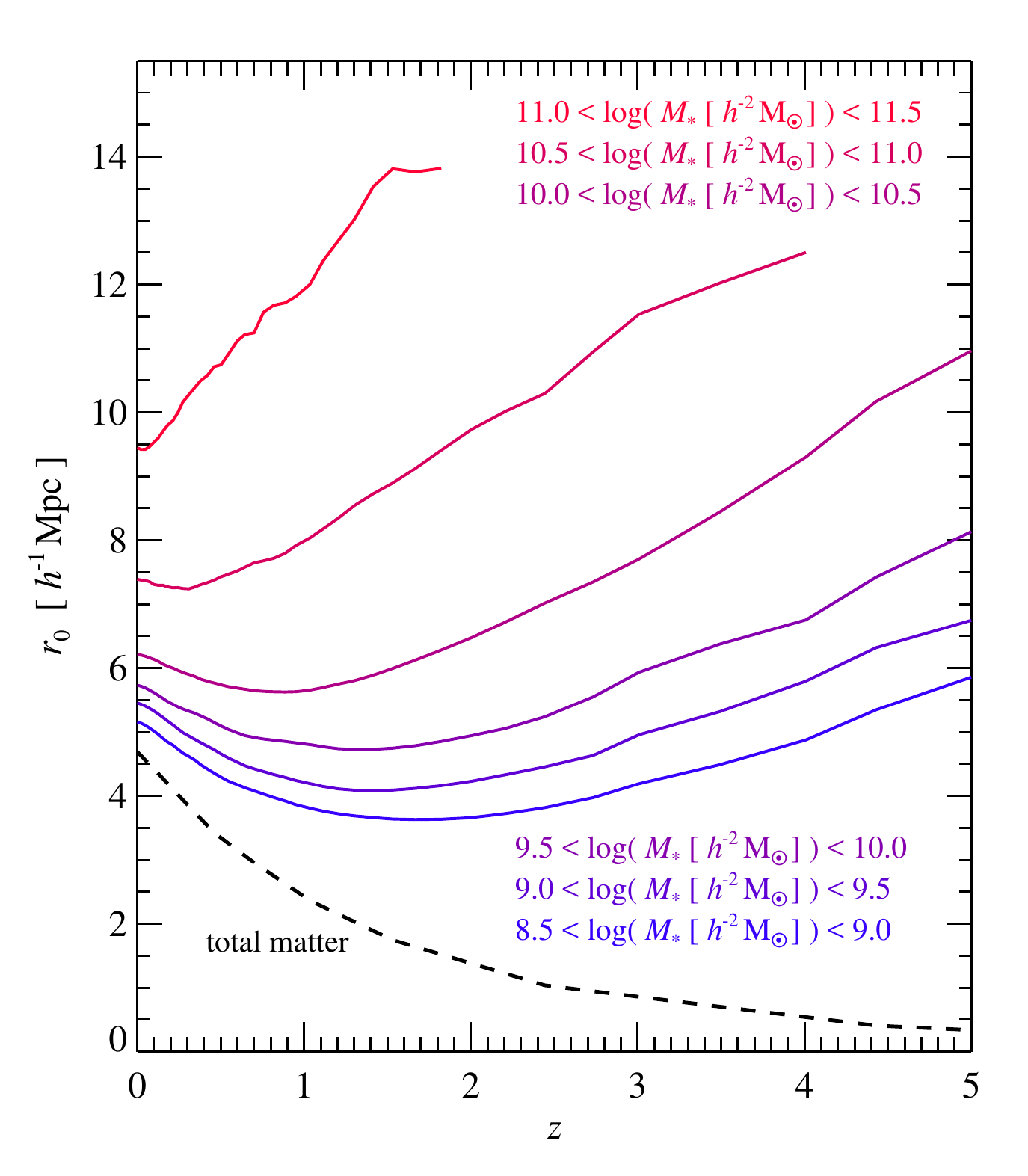}}%
  \resizebox{9cm}{!}{\includegraphics{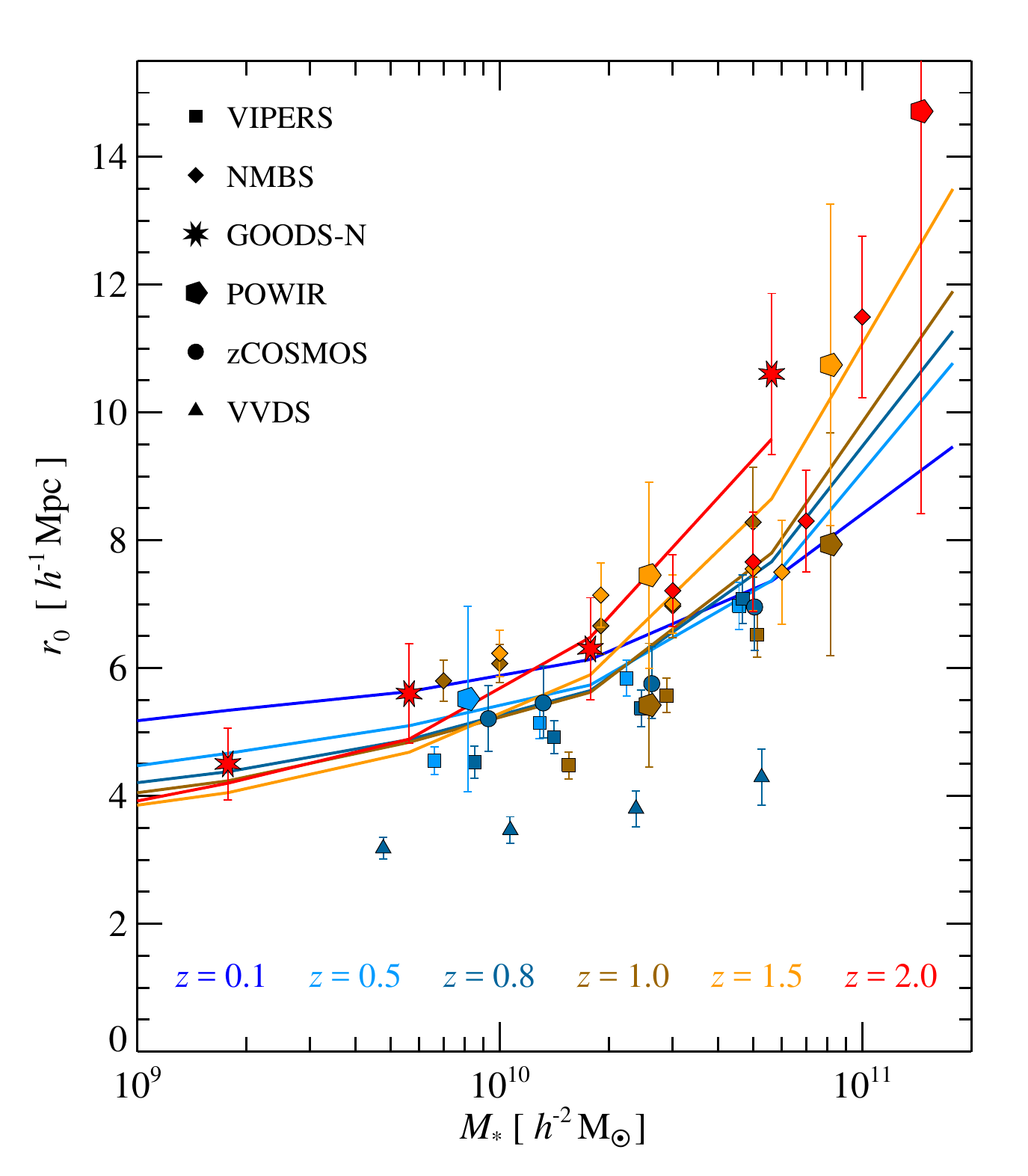}}
  \vspace*{-0.5cm} \caption{{\em Left panel:} Clustering length for
    different galaxy samples from the TNG300 simulation as a function
    of redshift. We show results for six different stellar mass ranges
    (coloured lines, as labelled). In each case, we define the
    clustering length as the scale where the real-space correlation
    function reaches unity, i.e.~$\xi(r_0)=1$. Deriving this through
    power-law fits to the real-space or projected correlation function
    over a range $5\,h^{-1} {\rm Mpc} < r < 20\,h^{-1}{\rm Mpc}$ gives
    very similar results. We also include the evolution of the
    correlation length of the total matter correlation function
    (dashed), which monotonically declines towards high redshift --
    quite unlike the galaxy samples which can be equally or even more
    strongly clustered at high redshift than today.  {\em Right
      panel:} Correlation length as a function of stellar mass for
    galaxies in TNG300 (solid lines), for different redshifts (colour
    key). We compare to data points from different observational
    surveys (symbols), in particular from VIPERS \citep{Marulli2013},
    NMBS \citep{Wake2011}, GOODS-N \citep{Lin2012}, POWIR
    \citep{Foucaud2010}, zCOSMOS \citep{Meneux2009} and VVDS
    \citep{Meneux2008}. The symbols have been coloured according to
    the characteristic redshift of the corresponding observational
    sample, showing weak systematic trends of clustering strength with
    redshift at fixed stellar mass.}
  \label{fig:clustlength}
\end{figure*}

In Figure~\ref{fig:2pcf-vvds}, we consider clustering at much higher
redshift of around $z\sim 1$, comparing different stellar mass samples
of TNG300 to results published for the VIMOS VLT Deep Survey
\citep[VVDS,][]{Meneux2008} and the DEEP2 galaxy redshift survey
\citep{Mostek2013}. The clustering signal of the simulated galaxies
agrees very well with DEEP2. However, while it appears close in shape
to VVDS, it is clearly somewhat stronger than this survey, a finding
similar to that reported by \citet{Meneux2008} for the Millennium
simulation \citep{Springel2005}, which shows a comparable excess. We
note however that VVDS may be affected by cosmic variance due to the
limited survey volume, and that the clustering lengths reported by the
survey lie low compared to other surveys at similar redshift.

We make this more explicit in Figure~\ref{fig:clustlength}, where we
show the correlation length $r_0$ (defined as $\xi(r_0)=1$) for
different stellar mass samples from TNG300 as a function of redshift
(left panel), or for samples at different redshift as a function of
stellar mass (right panel). The latter is also compared to results
reported for various galaxy surveys. Clearly, the clustering strength of
the simulated galaxies is a strong function of stellar mass at any
redshift. We note that the analysis of small-scale clustering of
\citet{Artale2016} in EAGLE did not find any clear evidence for an
increase of clustering strength with stellar mass or $r$-band
luminosity, in contrast to what we obtain here.

We also find that for a given stellar mass, the clustering length is a
function of redshift, but depending on stellar mass, the evolution
with redshift is not necessarily monotonic. For intermediate stellar
masses, the clustering length first declines towards high redshift and
then increases again, whereas for samples of very massive galaxies, it
only increases towards higher redshift. The correlation length of the
total matter, also included in Fig.~\ref{fig:clustlength} (left
panel), behaves very differently and monotonically declines towards
high redshift. Evidently, the bias between galaxies and matter is thus
a strong function of redshift; it is generally high at early times,
and then comes down and approaches values of order unity towards the
present epoch. 

It is interesting to compare various observational results for the
clustering length to these simulation predictions, as we do in the
right panel of Fig.~\ref{fig:clustlength}.  We consider data from VVDS
\citep{Meneux2008}, the Palomar Observatory Wide-field Infrared Survey
\citep[POWIR][]{Foucaud2010}, VIPERS \citep{Marulli2013}, the NEWFIRM
Medium Band Survey \citep[NMBS][]{Wake2011}, GOODS-N \citep{Lin2012},
and zCOSMOS \citep{Meneux2009}.  At a given stellar mass, our
simulations predict only weak variations of clustering strength with
redshift. For luminous galaxies with stellar mass above
$10^{10}\,{\rm M}_\odot$, the clustering strength increases towards
higher redshift, a trend that is reversed for lower mass galaxies at
low redshift.  We find that our simulation predictions are broadly
consistent with the data, which itself shows relatively large scatter,
precluding any strong conclusions at this point about whether these
subtle trends are also seen in the data.  The theoretical results
obtained here certainly provide strong motivation to start using the
evolution of clustering length for specific galaxy samples as an
important test of galaxy formation models.

\begin{figure}
  \centering
  \resizebox{8cm}{!}{\includegraphics{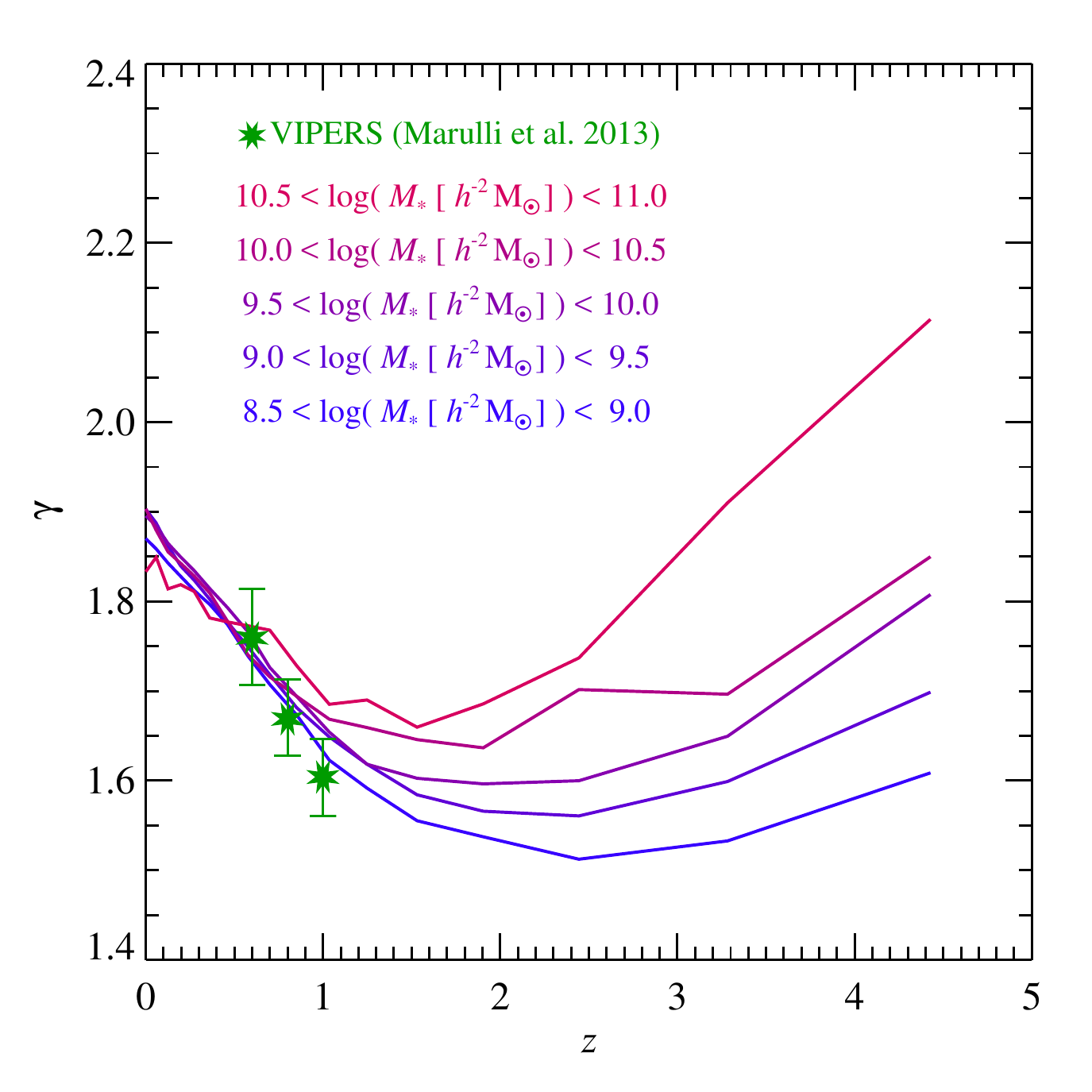}}
  \caption{Slope of the two-point galaxy correlation function as a
    function of redshift and stellar mass, for the same simulated
    galaxy samples of TNG300 considered in the left panel of
    Fig.~\ref{fig:clustlength}. The data points give results from the
    VIPERS survey \citep{Marulli2013} for
    $\log[M_\star / (h^{-2}M_\odot)] \simeq 10.15$, which agrees quite
    well with the redshift evolution we find here, and also has not
    found evidence for a significant stellar mass dependence of the
    slope in the low redshift regime. }
  \label{fig:clustslope}
\end{figure}

Many observational studies fit power-laws to the projected or
real-space correlation functions in order to infer the correlation
lengths and to represent the results in compact form. This is
motivated by the close to power-law shape of the galaxy
auto-correlation function at low redshift. In
Figure~\ref{fig:clustslope}, we show the power-law slope as a function
of redshift obtained by fitting each of our measured galaxy
correlation functions for different stellar mass samples, and at
different redshifts, over the range $1 < r / (h^{-1}{\rm Mpc}) <
15$. Strikingly, there is rather little dependence of the slope on stellar
mass, at least at low redshift. The slope is $\gamma \sim 1.6$ at
redshift $z\simeq 1$, and then steepens to $\gamma\simeq 1.8$ at
$z=0$. At intermediate redshifts, the low-mass stellar mass samples
show somewhat shallower slopes than the galaxy samples with higher
stellar mass, a difference that progressively becomes larger as they
steepen again towards high redshift. When compared to observations,
such as the VIPERS survey analysed in \citet{Marulli2013}, we see that
they show a very similar dependence of clustering slope on
redshift. Interestingly this survey also failed to detect a
significant stellar mass dependence of the slope for fixed redshift,
which is quite consistent with our results.

\section{Large-Scale Halo clustering}      \label{sec_haloes}

We now turn to an analysis of halo clustering, which is a central
concept in empirical models for galaxy large-scale structure, such as
HOD models, or more recently in SHAM models.  It is generally believed
that galaxies inherit the large-scale bias of their host halo, hence
understanding halo bias is often used as a way to sidestep the issue
of addressing galaxy bias directly. Recently, \citet{Jose2016}
formulated a model for the scale-dependence of halo bias which offers
the prospect to also extend this to quasi-linear scales.

\begin{figure}
  \centering
  \resizebox{8.5cm}{!}{\includegraphics{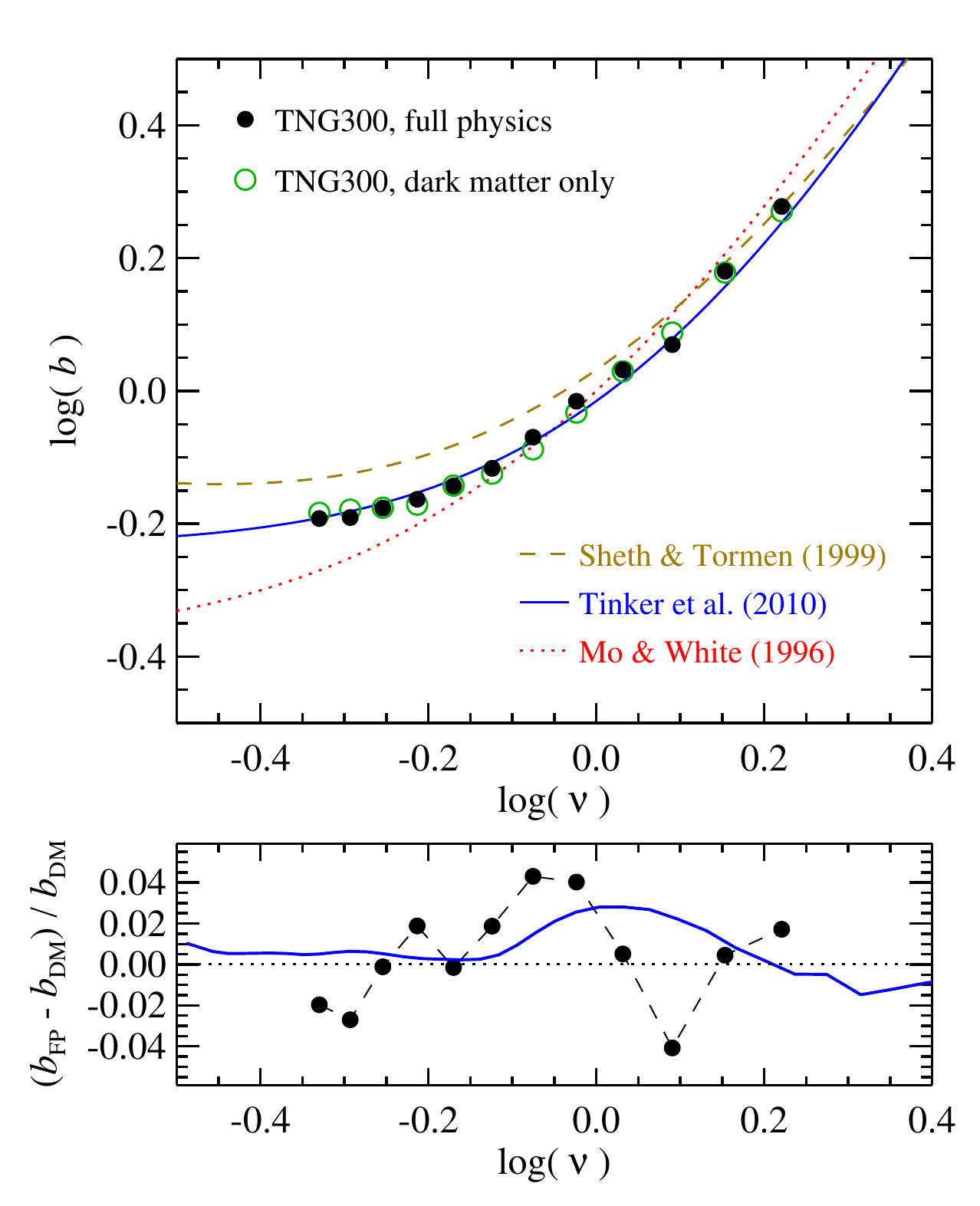}}
  \caption{{\em Upper panel:} Linear halo bias on large scales as a
    function of associated peak height $\nu$. We show results for the
    full physics (filled circles) and the dark matter only (empty
    circles) simulations of TNG300, and compare to the analytic models
    of \citet{Mo1996} and \citet{Sheth1999}, as well as to the
    empirical fit of \citet{Tinker2010} calibrated on a suite of
    collisionless N-body simulations. The latter describes our results
    quite well. {\em Lower panel:} Relative difference in halo bias
    between the full physics and dark matter only simulations, showing
    a scale-dependent variation of up to 3\%. This systematic
    difference is of the same order as the one expected (solid line)
    from the mass change of haloes due to baryonic effects, see
    Fig.~\ref{fig:masschange}.
    \label{fig:biashaloes_peakheight}}
\end{figure}

\begin{figure}
  \centering
  \resizebox{8.5cm}{!}{\includegraphics{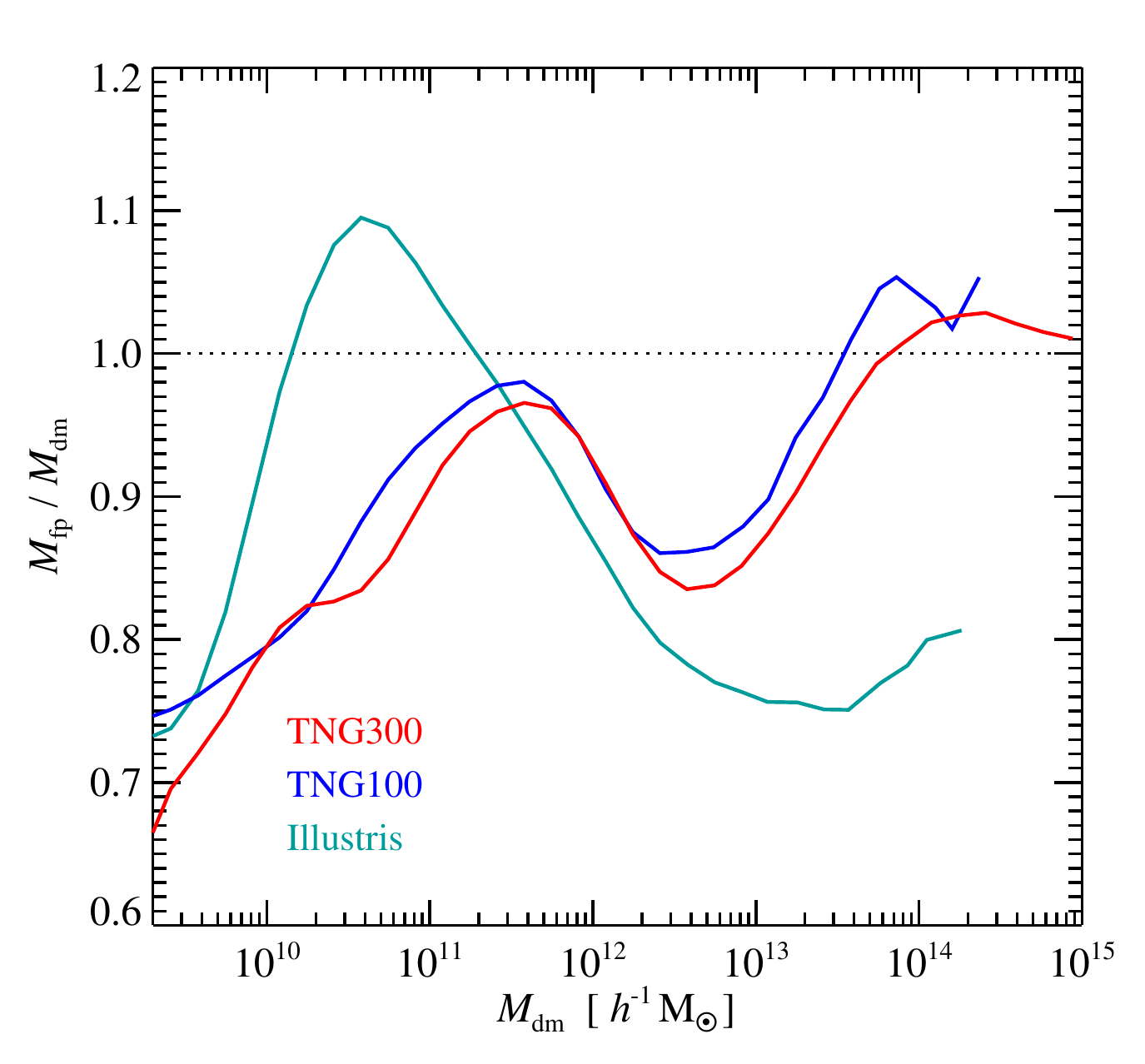}}
  \caption{Baryonic physics impact on the virial mass ($M_{200}$) of
    haloes, as a function of the mass of haloes in the
    corresponding dark matter only simulation. We show results
    obtained by abundance matching (i.e.~rank ordering the haloes by
    decreasing mass, and then comparing them in this order) for the
    halo populations in TNG300, TNG100, and, for comparison, the
    Illustris simulation \citep{Vogelsberger2014MNRAS}.}
  \label{fig:masschange}
\end{figure}

It has long been realised that dark matter haloes more massive than the
characteristic halo mass are positively biased with respect to the
matter, and less massive haloes exhibit a negative bias. This can be
understood based on the clustering of peaks in Gaussian random fields.
Similar to derivations of the halo mass function, this gives rise
to analytic models for halo bias. This is concisely expressed in the
model of \citet{Mo1996}, which determines the halo bias in terms of
peak height, making the theoretical prediction in principle universal
and independent of redshift. Many N-body simulations have been used to
test this prediction, generally finding that it works quite well, but
with some residual discrepancies that motivate the development of
improved models \citep[e.g.][]{Sheth1999, Pillepich2010, Tinker2010}.

The situation is reminiscent of the halo mass function, where the
basic spherically symmetric formulation by \citet{Press1974} provides
a decent first order approximation that can be substantially improved
by models of ellipsoidal collapse \citep{Sheth2001}. However, the
latter still shows some discrepancies compared to N-body simulations,
which can be addressed through empirical fitting functions to the
numerical results \citep[e.g.][]{Jenkins2001}.

In Figure~\ref{fig:biashaloes_peakheight}, we show the linear halo
bias on the largest scales, as a function of peak height, in the form
it has been analysed in a large number of cold dark matter simulation
studies. We include results for the full physics run of TNG300 (filled
circles), and for the dark matter only version of the same model (open
circles). For comparison, we also show the predictions of
\citet{Mo1996}, \citet{Sheth1999}, and \citet{Tinker2010}. The latter
provides clearly a very good fit to our results. The differences
between the full physics simulation and the dark matter only results
are displayed in the lower panel of
Fig.~\ref{fig:biashaloes_peakheight}, and are quite small. There are
some systematic distortions in the halo bias of up to 3\% induced by
baryonic physics. We argue that these changes can be largely explained
by a modification of the halo masses themselves. For example, when
feedback effects expel gas from a halo, its mass is lowered and its
subsequent non-linear growth can be slowed as well. The associated
initial density fluctuation peak is not changed by this,
however. Knowing this halo mass change quantitatively (which we
determine below), we can predict the expected change of halo bias as a
function of peak height, based on the model of \citet{Tinker2010}. The
result of this is shown as a solid line in the lower panel of
Figure~\ref{fig:biashaloes_peakheight}. While this does not reproduce
the (somewhat noisy) measurements in detail, the predicted effect is
of very similar magnitude to the one measured, suggesting that this is
indeed the dominating effect.

This estimate has made use of the results of
Figure~\ref{fig:masschange}, where we show the mass change of haloes
due to the inclusion of baryonic physics. To this end we compare the
$M_{200}$ masses of haloes at equal comoving abundance in
corresponding pairs of full physics and dark matter only
simulations. This is simply achieved by sorting the haloes by mass in
descending order, and then comparing them at equal rank, a procedure
that yields very similar (albeit not identical) outcomes to
cross-matching haloes by the particle-IDs of their dark matter
content.  We include results for TNG300, TNG100 and the Illustris
simulation. {In TNG, a particularly strong impact of baryonic
  physics occurs for halo masses around
  $10^{13}\,h^{-1}{\rm M}_\odot$, which is due to the comparatively
  sudden onset of strong kinetic-mode AGN feedback, as can be verified
  through the increase of the associated energy input in galaxies of
  the associated stellar mass \citep{Weinberger2017b}.} However,
consistent with the reduced impact of baryonic physics on the power
spectra, we find that the TNG model shows overall a much weaker impact
on halo masses than the Illustris feedback model \citep[see
also][]{Vogelsberger2014}. In particular, the suppression of halo
masses due to AGN feedback sets in at higher masses, and is restricted
to a narrower mass range, with poor clusters of galaxies already being
largely unaffected.  On the other hand, in the halo mass range
$10^{10}- 10^{11}\,h^{-1}{\rm M_\odot}$, the TNG model shows a
stronger effect on its halo masses than Illustris, reflecting its
modified wind model. {Looking also at the results for
other feedback implementations \citep{Velliscig2014}, is therefore clear that 
the variation of halos masses depends
quite sensitively on the employed feedback model.}

\section{Scale-dependent Bias}        \label{sec_bias}

\begin{figure}
  \centering
  \includegraphics[width=0.47\textwidth]{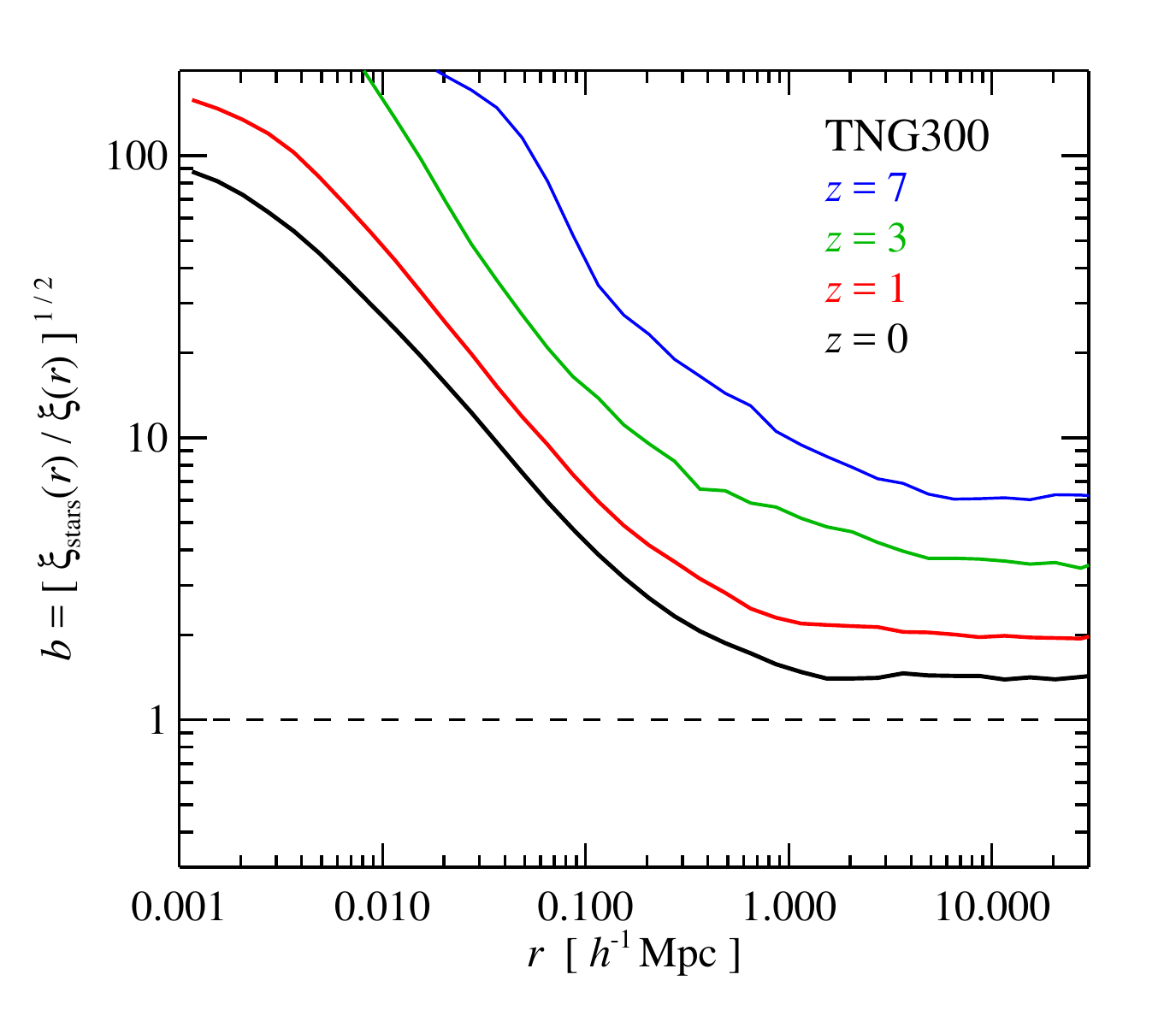}
  \caption{Bias of the stellar mass distribution relative to the total
    matter in TNG300 at different redshifts. At $z=0$, the bias of
    the stellar mass becomes independent on distance for scales larger than
    $\sim 1\,h^{-1}{\rm Mpc}$, but it is still non-zero. At earlier
    times, the bias is much larger, and scale-dependent effects are
    seen out to larger scales.}
  \label{fig:matter2cf_bias}
\end{figure}

In Figure~\ref{fig:matter2cf_bias}, we consider the bias of all the
stellar mass in our TNG300 simulation relative to the total matter,
here in terms of the real-space clustering. This corresponds to
$b(r)=[\xi_\star (r) / \xi(r)]^{1/2}$ for the results in
Fig.~\ref{fig:matter2cf}. We clearly see the very large positive bias
of about $b\sim 7$ at the highest examined redshift, which then
progressively comes down towards the present epoch. At $z=0$, the bias of
the stellar mass is still positive with a value of about
$b\simeq 1.4$. Interestingly, the scale-dependence of this bias sets in
earlier (i.e. on larger scales) at high redshift than at low redshift.

A scale-dependent bias can be a great challenge for the interpretation
of galaxy redshift surveys. Such a scale-dependence will naturally
arise from mild quasi-linear and fully non-linear evolution, but even
when the bias is considered relative to the non-linearly evolved
density field, it is not clear a priori up to which scales galaxies
can be used as faithful tracers for the mass distribution by simply
invoking the value of the linear bias on the largest scales. Another
complication is that the bias is expected to strongly depend on the
sample selection procedure. Tracers with the same number density but
of different type can exhibit substantially different biases, and can
also be affected to different degrees by scale-dependence.

\begin{figure}
  \centering
  \includegraphics[width=0.47\textwidth]{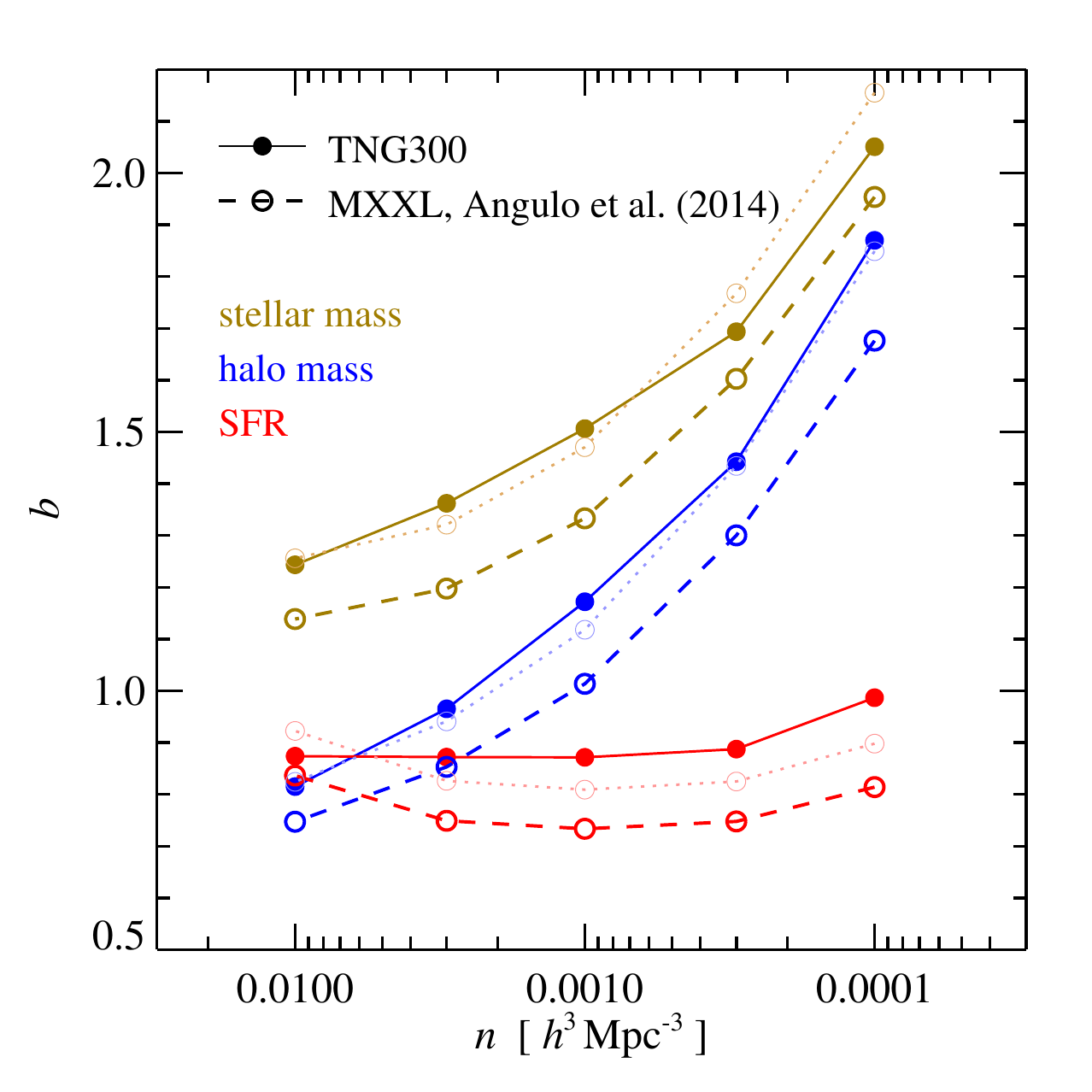}
  \caption{Large-scale halo and galaxy bias as a function of tracer
    number density for samples selected by stellar mass, halo mass, or
    instantaneous star formation rate. We compare our results for
    TNG300 with an equivalent analysis by \citet{Angulo2014} for a
    semi-analytic model applied to the Millennium-XXL simulation
    (dashed). Our bias values show very similar trends but lie
    consistently higher. When we correct the dark matter correlation
    function of MXXL for its higher normalisation ($\sigma_8=0.9$)
    relative to TNG ($\sigma_8=0.8159$), this difference goes largely
    away (dotted lines).}
  \label{fig:comp_to_mxxl}
\end{figure}

In Figure~\ref{fig:comp_to_mxxl}, we demonstrate the dependence on
tracer type explicitly by showing the linear bias on large scales as a
function of tracer number density, for three different selection
criteria. In particular, we consider haloes according to their virial
mass ($M_{200}$), galaxies selected by their stellar mass ($M_\star$),
and galaxies selected by their instantaneous star formation rate
($\dot M_\star$). In each case, the objects are sorted in descending
order and included top down until the corresponding space density
is reached. Remarkably, the galaxies selected according to their star
formation rates show only a very weak variation of their bias (which
is in fact an anti-bias) with tracer density. This shows that these
galaxies do not tend to populate the most massive haloes -- which
makes sense because these haloes are often quenched and hence are not
natural hosts of star-forming galaxies. In contrast, galaxies selected
by stellar mass show a large positive bias that increases strongly
towards more luminous systems. This reflects the trend of higher bias
values for rarer and hence more massive haloes.

We have compared our results in Fig.~\ref{fig:comp_to_mxxl} to a
similar analysis carried out by \citet{Angulo2014} for the
Millennium-XXL simulation \citep{Angulo2012} on the basis of a
semi-analytic galaxy formation model. There is good qualitative
agreement, but our bias values are systematically higher. Most of the
difference can simply be explained by the higher value of
$\sigma_8=0.9$ adopted in \citet{Angulo2014}. As the clustering
pattern of a given tracer evolves comparatively little in time,
whereas the dark matter auto-correlation on large scales grows
according to linear theory, we can to first order correct for this by
comparing to an earlier output of Millennium-XXL when its
normalization corresponds to our value of $\sigma_8=0.8159$. Or
simpler still, we can adjust their bias results by a factor of
$0.9/0.8159$, thus effectively bringing the MXXL's dark matter
correlation function to the less evolved state corresponding to our
simulation.  This yields the dotted results in
Fig.~\ref{fig:comp_to_mxxl}, which are in good agreement with TNG300.

\begin{figure*}
  \centering
\includegraphics[width=0.5\textwidth]{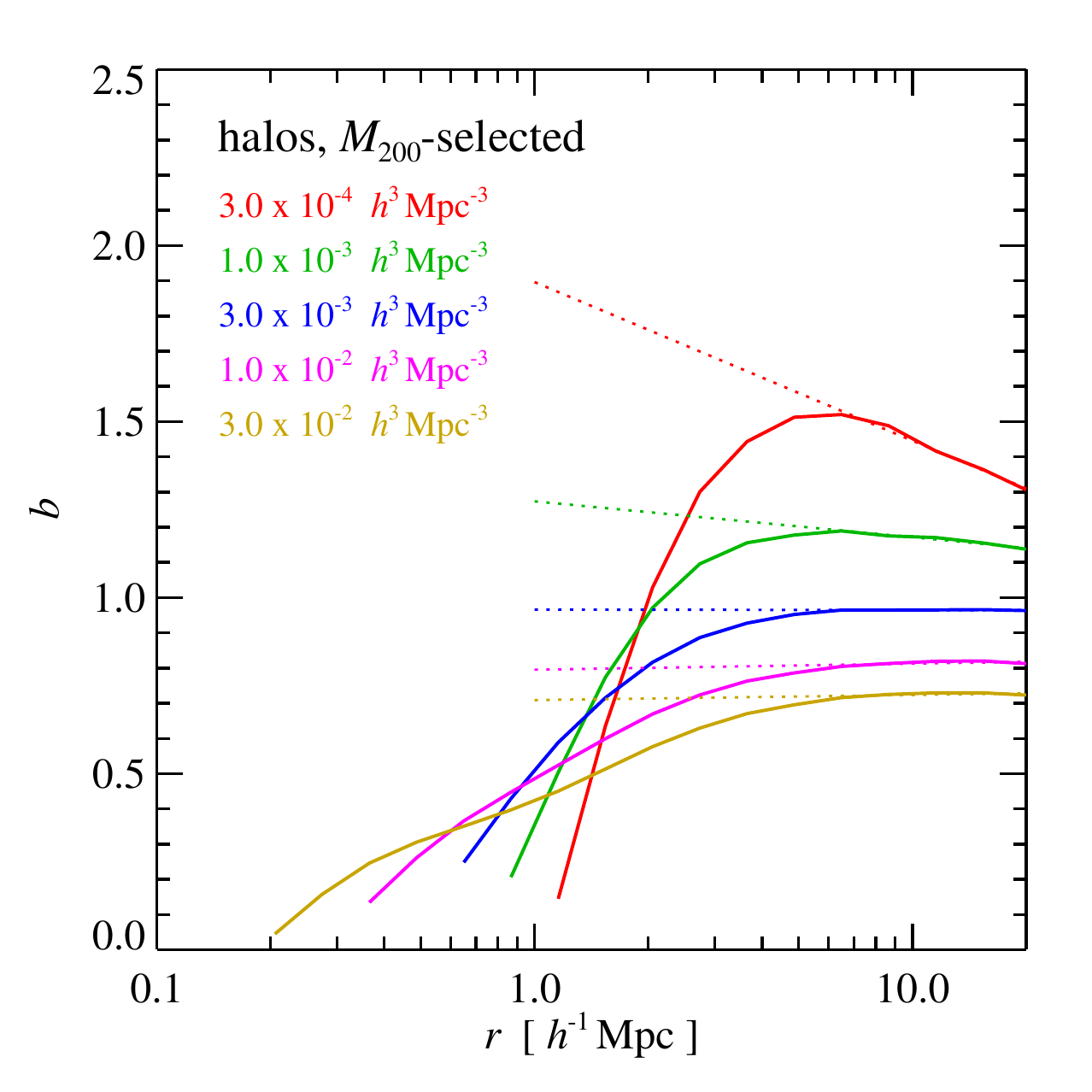}%
\includegraphics[width=0.5\textwidth]{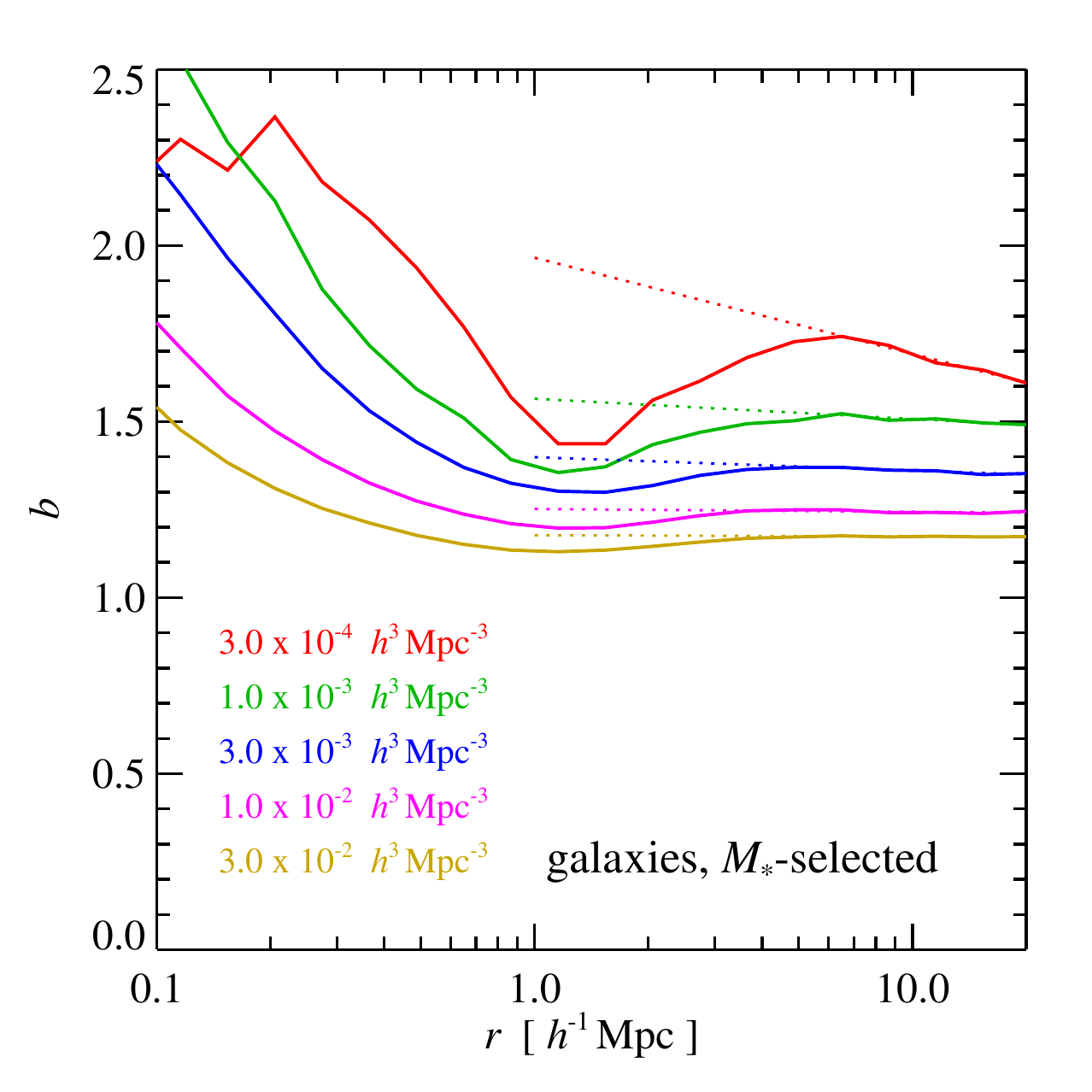}\\
\includegraphics[width=0.5\textwidth]{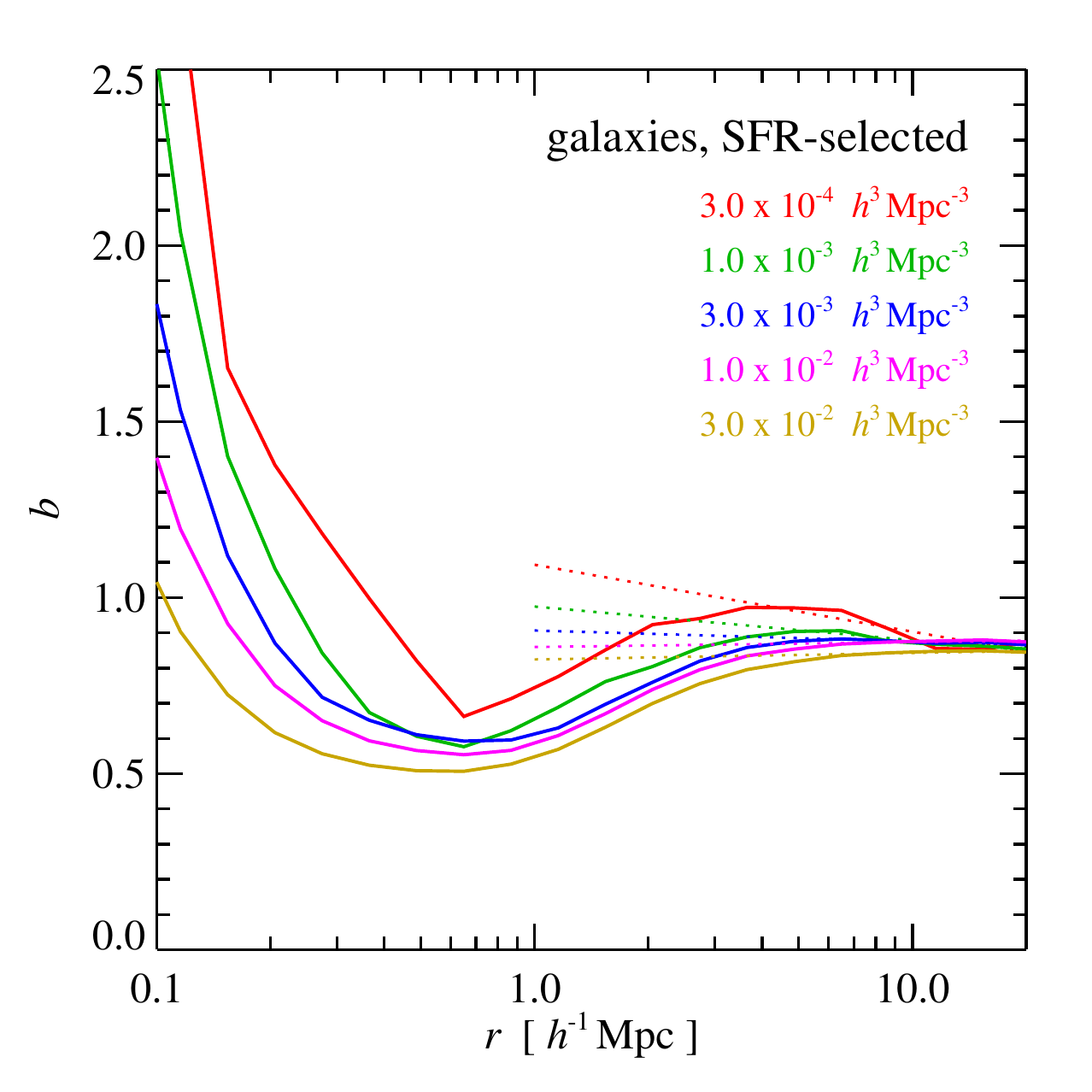}%
\includegraphics[width=0.5\textwidth]{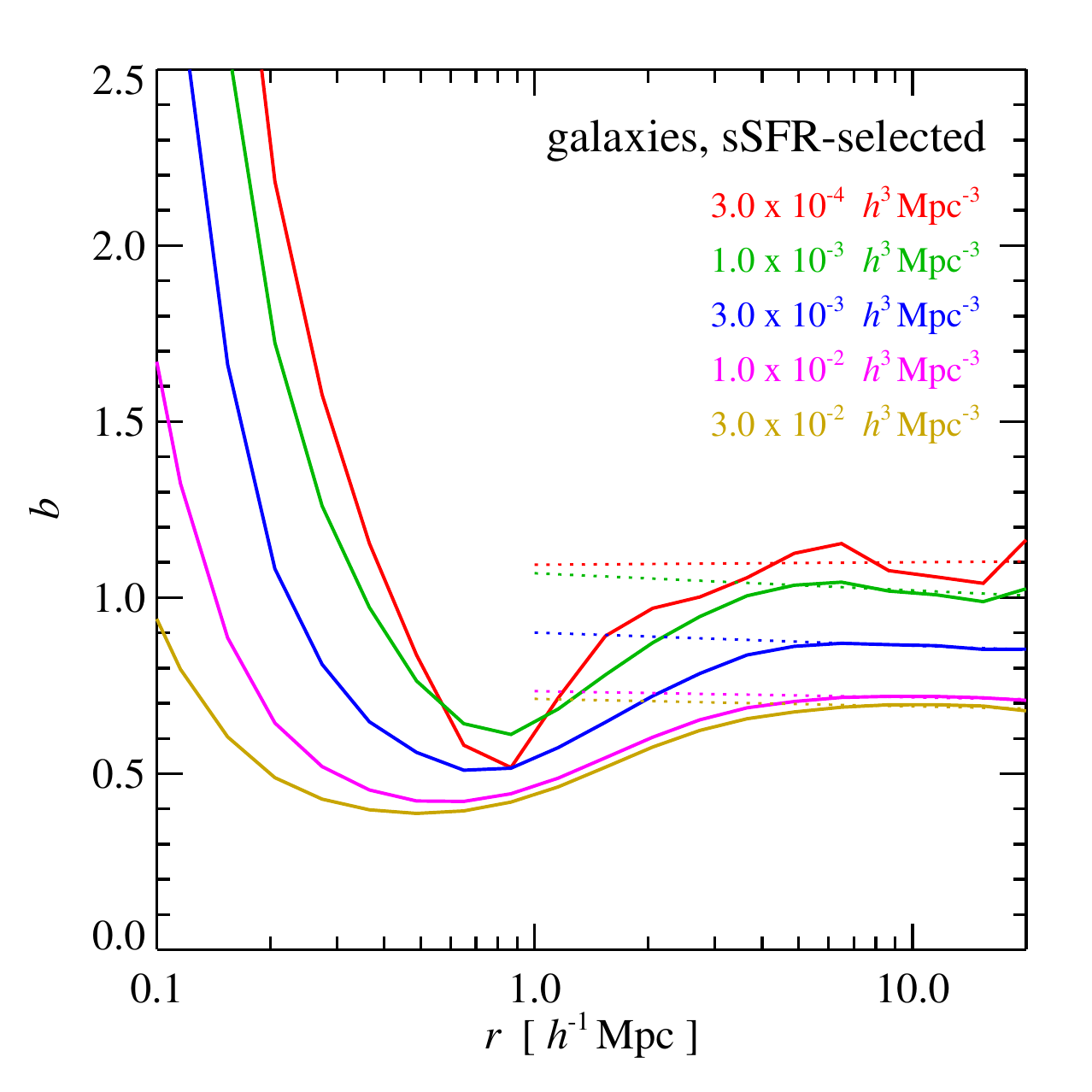}\\
\caption{Scale-dependent bias (based on real-space correlation
  functions, $b(r) = [\xi_{\rm tracer}(r)/\xi_{\rm dm}(r)]^{1/2}$) for
  different space-densities of tracer objects, selected in a variety
  of ways. The top-left panel shows haloes, with the most massive ones
  selected according to their $M_{200}$-mass up to a certain space
  density, as labelled. The top-right panel selects galaxies according
  to their stellar mass ($M_\star$), the bottom left according to
  their instantaneous star formation rate ($\dot M_\star$), and the
  bottom right according to their specific star formation rate
  ($\dot M_\star / M_\star$). All results are given at $z=0$ for the
  TNG300 simulation. Dotted lines are fits to the region
  $5\,h^{-1} {\rm Mpc} < r < 25\,h^{-1} {\rm Mpc}$ and are meant to
  guide the eye only, illustrating the tentative evidence for
  significant scale-dependencies of the bias over this region in some
  of the samples. }
  \label{fig:bias_r_dependence}
\end{figure*}

Instead of just looking at the linear bias on the largest scales, it
is much more interesting to consider the bias of different tracers
also as a function of scale. We do this in
Figure~\ref{fig:bias_r_dependence} for the real-space two-point
correlation function.  Again, we consider tracers with different space
densities, with five values ranging from $3 \times 10^{-4}$ to
$3 \times 10^{-2}\,h^3\,{\rm Mpc}^{-3}$, using a selection by halo
mass, galaxy stellar mass, instantaneous star formation rate, and
current specific star formation rate ($\dot M_\star/ M_\star$). For
our TNG300 volume, the samples then contain 2585 objects at a space
density of $3\times 10^{-4}\,h^3\,{\rm Mpc}^{-3}$, and 258454 at the
highest considered density of $3\times 10^{-2}\,h^3\,{\rm Mpc}^{-3}$.
We note that these space densities cover the range considered in a
number of ongoing or planned large galaxy surveys that target
cosmology, hence we expect effects of similar magnitude in real galaxy
surveys. For definiteness, for the halo samples the limiting $M_{200}$
values of the five considered space densities are
$9.81 \times 10^{12}$, $2.91\times 10^{12}$, $1.02 \times 10^{12}$,
$2.86 \times 10^{11}$, and $7.85\times 10^{10}\,h^{-1}{\rm M}_\odot$,
respectively.  For the stellar mass samples, the limiting values are
$6.84\times 10^{10}$, $3.08\times 10^{10}$, $1.54\times 10^{10}$,
$4.39 \times 10^9$, and $4.66\times 10^8\,h^{-2}{\rm M}_\odot$.  The
selection according to star formation rate corresponds to limiting
values of $5.25$, $3.03$, $1.55$, $0.468$ and
$0.066\,{\rm M}_\odot {\rm yr}^{-1}$. Finally, the specific star
formation rate selection is based on cut-off values of $10.28$,
$4.43$, $1.27$, $0.492$ and $0.212\,h\,{\rm Gyr}^{-1}$.

For the halo samples in Fig.~\ref{fig:bias_r_dependence}, the bias of
the two-point correlation functions shows a clear short-range
exclusion effect, with the bias function suddenly dropping
precipitously and rapidly towards short distances. Halo samples
dominated by relatively low mass haloes show no significant scale
dependence for $r > 7\,h^{-1}{\rm Mpc}$, but more massive haloes do. A
similar behaviour is found for the stellar mass samples, except that
on scales of $r \sim 1\,h^{-1}{\rm Mpc}$ a mild decrease of the bias
is seen, followed by a strong rise towards small scales. The bias at a
given space density for samples selected by stellar mass is always
much higher than for haloes, and also when galaxies are selected by star
formation rate or specific star formation rate. For the latter two
samples, the dip in bias at intermediate scales and the small-scale
rise are much more pronounced than for halo or stellar mass samples.

We now turn to the question of whether such scale dependent biases
also affect the baryonic acoustic oscillations (BAOs).  These are an
important cosmological resource, and in particular a primary
observational handle to constrain the cosmic expansion history and
thus models of dark energy. Detecting the BAOs not just in the cosmic
microwave background but also in galaxy surveys or quasar absorption
line studies at intervening redshifts is therefore a major
goal in observational cosmology \citep[e.g.][]{Eisenstein2005,
  Cole2005, Wang2017}.

\begin{figure*}
  \centering
  \includegraphics[width=0.49\textwidth]{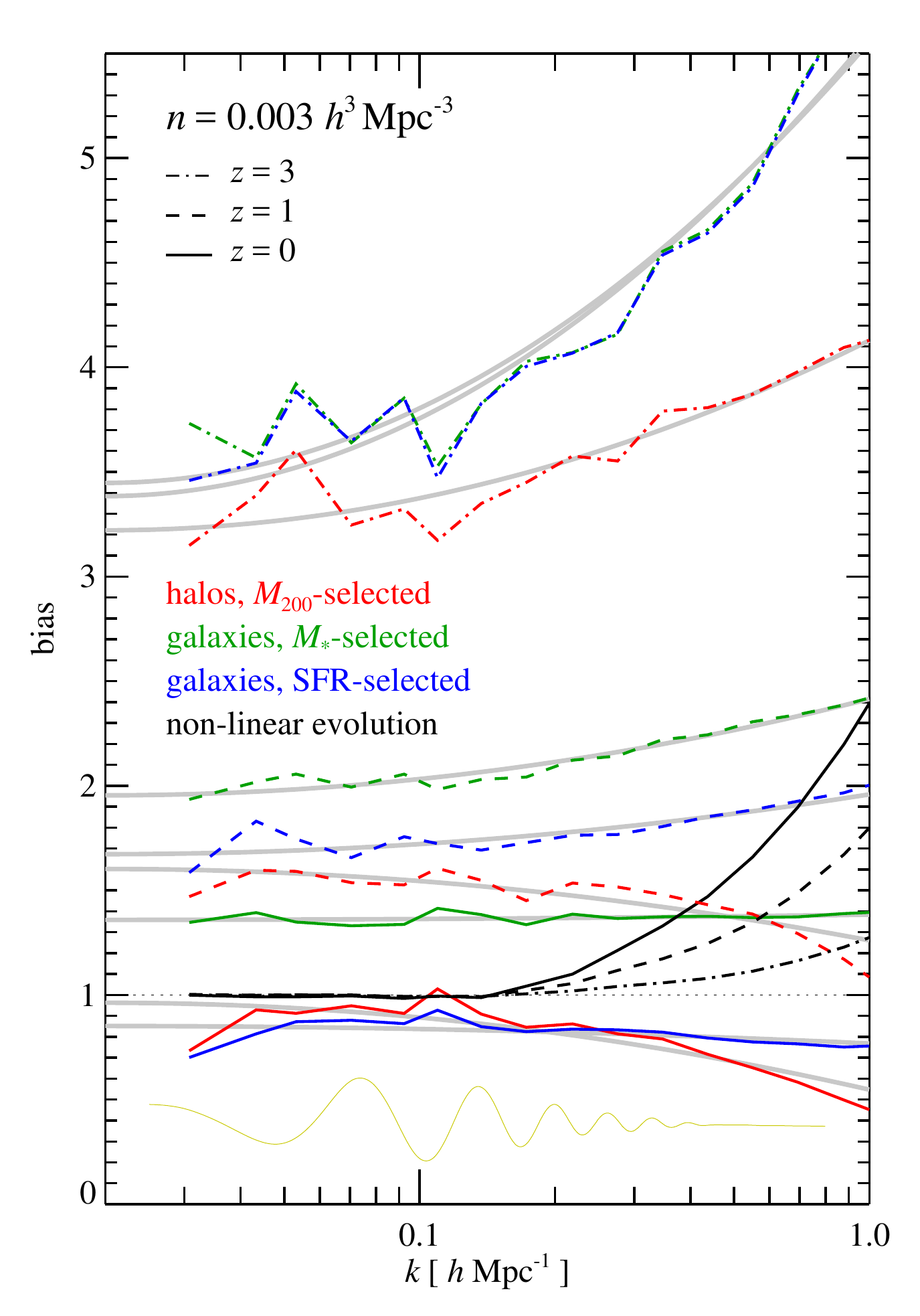}%
  \includegraphics[width=0.49\textwidth]{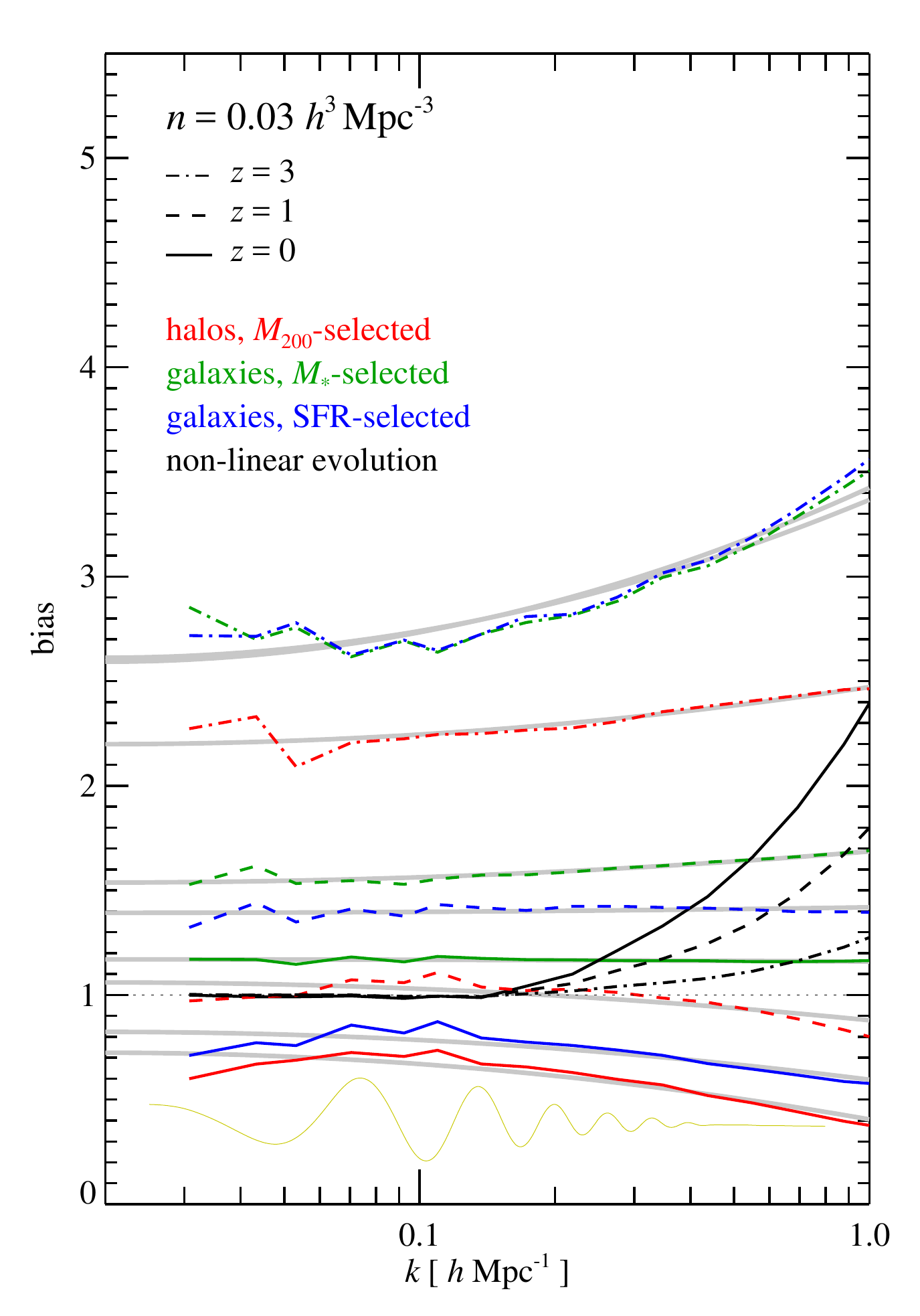}
  \caption{Bias as a function of wavenumber $k$ (based on power spectrum ratios, i.e.~$b(k) = [P_{\rm tracer}(k)/P_{\rm matter}(k)]^{1/2}$) for different types of tracers and different space densities, for the TNG300 simulation at $z=0$. The shown samples are
    a subset of those examined 
   in Fig.~\ref{fig:bias_r_dependence} in terms of real-space
    clustering, and represent haloes chosen by virial mass (red), galaxies
    selected by stellar mass (green), and galaxies selected by star formation
    rate (blue). Different redshifts are shown as solid, dashed and dot-dashed lines, respectively, with the left and right panels giving results for different tracer densities. For comparison, we also show the effective ``bias'' due to non-linear evolution of the matter density power spectrum relative to the linear theory prediction for the corresponding redshifts (black lines). The faint yellow lines illustrate the location of the baryonic acoustic oscillations. 
{Thick grey lines denote best fits according to the scale-dependent bias formula of equation~(\ref{eqnfit}) and best-fit parameters listed in Table~\ref{TabFits}.}
    \label{fig:bias}}
\end{figure*}

Our TNG300 simulation box is just large enough to see the baryonic
acoustic oscillations (BAO) in the total matter power spectrum, but
the number of available large-scale modes over the relevant range is
too small to directly measure the oscillations with the required
accuracy for cosmological inferences. However, we can still measure
the power spectra of different tracers and determine their bias
relative to the non-linear matter power spectrum. Taking the ratio of
the two power spectra eliminates much of the cosmic variance
fluctuations due to the specific realisation of our large-scale
modes. The result is seen in Figure~\ref{fig:bias}, where we show our
bias measurements for two different space densities, $3\times 10^{-3}$
and $3\times 10^{-2}\,h^3\,{\rm Mpc}^{-3}$, and for tracers selected
by halo virial mass, galaxy stellar mass, and galaxy star formation
rate.  We include results for redshifts $z=3$, $z=1$, and $z=0$.

Again, it is evident from the results that the bias depends strongly
on tracer type, redshift, and space density. Interestingly, over the
$k$-range of the baryonic acoustic oscillations (indicated as thin yellow
lines in Fig.~\ref{fig:bias}), clear variations of the measured bias
values are detected. To quantify this scale dependence over the range
$0.02 < k / (h\,{\rm Mpc}^{-1})< 1.0$, we fit our measurements with a
very simple scale-dependent bias model of the form
\begin{equation}
b(k) = b_0 + \beta \left(\ln\frac{k}{k_0}\right)^2. 
\label{eqnfit}
\end{equation}
Here $b_0$ represents the large-scale linear bias, while $\beta$
measures the strength of the scale dependence. We set
$k_0=0.02\,h\,{\rm Mpc}^{-1}$ in our fits so that
${\rm d}b(k)/{{\rm d}k} = 0$ at $k = k_0$. The resulting smooth bias
laws $b(k)$ are shown with thick grey lines in the figure, {and
  the corresponding fit parameters are given in Table~\ref{TabFits}}.  The very
good statistics we have for the full matter distribution also allows
us to reliably measure the ratio $[P(k)/P_{\rm lin}(k)]^{1/2}$ of the
full matter power spectrum relative to the linear theory power
spectrum. This is shown as black lines in Fig.~\ref{fig:bias} for
redshifts $z=3$, $z=1$, and $z=0$, and can be interpreted as an
effective bias that encodes the non-linear clustering evolution. The
bias of a tracer sample relative to the linear theory power spectrum
is then the product of this effective bias with the intrinsic bias of
the tracers.

\begin{figure}
  \centering
  \includegraphics[width=0.49\textwidth]{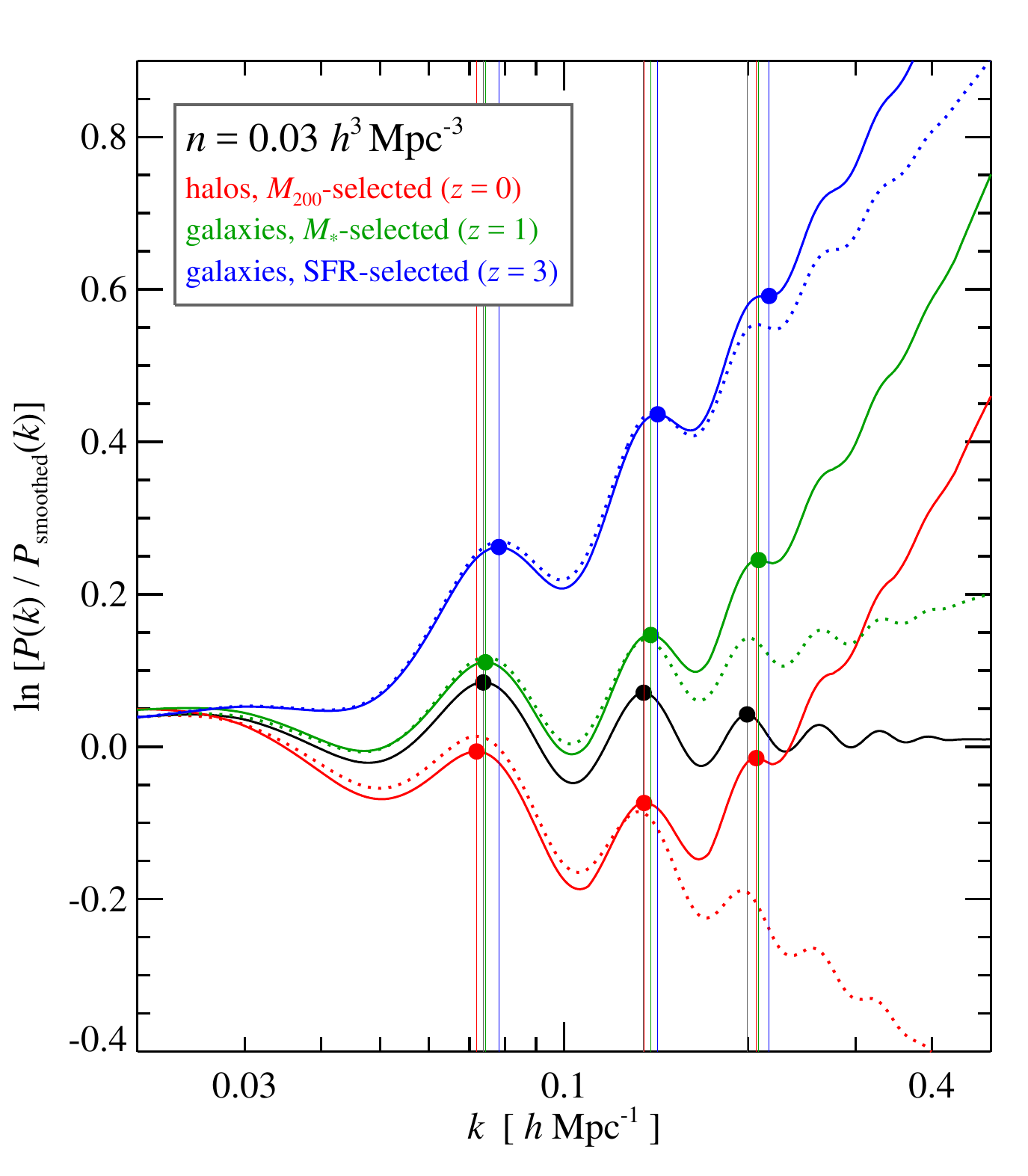}
  \caption{Power spectra estimates relative to the linearly evolved,
    smoothed initial power spectrum at different times, over the
    $k$-range containing the baryonic acoustic oscillations.  The
    dotted lines refer to power spectra obtained by multiplying the
    linear theory power spectrum by the fitted $b^2(k)$-factors of
    different tracers, which are shown in Fig.~\ref{fig:bias}. The solid lines
    additionally take effects of non-linear evolution into account, at
    the level of the change of the mean power per mode, as
    encapsulated by the black lines in Fig.~\ref{fig:bias}. Note that this
    neglects the damping and broadening of the wiggles due to mode-mode
    coupling. Three combinations of tracer type and redshift are shown with different
    colours, as labelled, with the relative bias on the largest scales
    renormalised to unity, and all for comoving tracer number density of $n=0.03\,h^{3}{\rm Mpc}^{-3}$. The unperturbed linear theory BAOs are
    shown as a black line. The positions of the local maxima in the
    first three wiggles have been located and marked
    with circles and thin vertical lines. The scale-dependent bias
    leads to sizeable shifts $\Delta k/k$ of up to 6\% in these peak positions, but thanks to the smooth variation of the bias with $k$, this distortion of the acoustic scale can be largely eliminated by template fitting of the expected BAO-signal.
    \label{fig:BAO}}
\end{figure}

We can use these results to obtain an estimate of the distortion of
the BAO features in the matter power spectrum due to non-linear
evolution and scale-dependent biases. This is shown in
Figure~\ref{fig:BAO}, where we modify the linear theory power spectrum
by the effective bias encoding non-linear evolution, and the
scale-dependent bias for three example tracers from
Fig.~\ref{fig:bias}. We show in each case the estimated evolved power
spectrum divided by a smoothed, wiggle-free linear theory power
spectrum, in order to bring out the BAOs more clearly. We stress that
this estimate only accounts for effects of non-linear evolution at the
level of the mean power per mode. In reality, mode-coupling effects
will tend to wash out and broaden the BAO features, and can even give
rise to additional small shifts in the peak positions. This is not
accounted for in our simple illustrative analysis shown in
Fig.~\ref{fig:BAO}, but this effect can be relatively well understood
for the dark matter density field based on renormalised perturbation
theory \citep{Crocce2008}.

Even though we have refrained from including in Fig.~\ref{fig:BAO}
the tracer samples with the strongest scale-dependent biases (which
occur for massive or star-forming galaxies at $z=3$) and neglected
mode-mode coupling, the distorting effect on the BAO features is
substantial. In particular, the $k$-positions of the local maxima of
the peaks are shifted by several percent, and depending on the precise
tracer or redshift considered, the shift can be both positive or
negative. Superficially, this may sound like very bad news for the
cosmological use of the BAO features. However, these shifts can be
corrected for by fitting the observed BAO features with the expected
signal template \citep[e.g.][]{Seo2008, Angulo2014, Prada2016}. For
example, one could use a simple model of the form
\begin{equation}
P_{\rm obs}(k) = (c_0 + c_1 k + c_2 k^2) P_{\rm lin}(k/\alpha) , 
\end{equation}
where $c_0$, $c_1$, $c_2$ and $\alpha$ are fit parameters. The $c_i$
describe a polynomial fit to the scale-dependent bias that is
empirically determined from the data, while $\alpha$ is a stretch
factor that is supposed to pick up a real shift of the acoustic scale,
if it exits.

Fitting such a model to the distorted BAOs of Figure~\ref{fig:BAO}
indeed recovers $\alpha$ values that are very close to unity: in the
case of the halo sample we obtain $\alpha-1 = -0.215\%$, for the
stellar mass selected galaxies $0.048\%$, and for the SFR-selected
sample $0.089\%$. The reason why this works so well is that the
scale-dependent bias we detect and the effects of non-linear evolution
vary smoothly with scale. When one knows what to look for, then they
can be taken out very well. And in our case, we have prescribed the
expected linear theory template precisely, without any measurement
errors and without a damping of the higher order wiggles by non-linear
evolution.

For real data, the conditions are not nearly as favourable, making the
possible systematic impact of scale-dependent bias on BAO studies
still an interesting research question.  At the very least, it is to
be expected that there will be a small impact on the constraining
power of observational surveys \citep[e.g.][]{Amendola2015}. The
ability of our simulation models to make accurate predictions for
galaxy bias should help precise characterisations of such
effects. This however requires a proper modelling of effects such as
mode-mode coupling, redshift space distortions, and observational
errors and selection effects, which is beyond the scope of this paper.

\section{Discussion and Conclusions}             \label{sec_conclusions}

In this study, we have analysed the matter and galaxy clustering in a
new suite of high-resolution hydrodynamical calculations of galaxy
formation, the IllustrisTNG simulations. Besides numerous improvements
in the treatment of feedback effects and in the numerical accuracy of
the simulations, an important advance of IllustrisTNG lies in its push
to higher volume, allowing a more faithful sampling of the halo and
galaxy distribution including rarer objects, and in particular, a much
better representation of cosmic large-scale structure. The latter is
probed in powerful ways by current and upcoming galaxy redshift
surveys (such as EUCLID, DES, or eBOSS). The precision with which
theoretical galaxy formation models can explain the clustering of
galaxies as a function of stellar mass, colour, star formation,
redshift, etc., offers the prospect to constrain and test such models
in interesting ways, as well as lifting possible modeling
degeneracies. At the same time, clustering data is a critical
component of modern cosmological constraints, including those that
seek to tie down the cosmic expansion history and inform about the
nature of dark energy. Here simulation predictions can inform about
possible observational biases and help to eliminate them.

\begin{table}
\begin{center}
\setlength{\tabcolsep}{4pt}
\begin{tabular}{cccrcrcr}
\hline
$n$ & \multirow{2}{*}{sample} & \multicolumn{2}{c}{$z=0$} & \multicolumn{2}{c}{$z=1$} & \multicolumn{2}{c}{$z=3$}\\
$[h^3{\rm Mpc}^{-3}]$ & & $b_0$ &  \multicolumn{1}{c}{$\beta$} & $b_0$ & \multicolumn{1}{c}{$\beta$} & $b_0$ & \multicolumn{1}{c}{$\beta$}\\
\hline
      & $M_{200}$  & $0.96$ & $-0.0271$ & $1.60$ & $-0.0223$ & $3.22$ & $0.0593$ \\
0.003 & $M_{\star}$ & $1.36$ & $0.0016$  & $1.95$  & $0.0301$ & $3.45$ & $0.1378$ \\
      & SFR       &  $0.85$ & $-0.0054$ & $1.67$ & $0.0187$ & $3.38$ & $0.1435$ \\
\hline
      & $M_{200}$  & $0.72$ & $-0.0208$ & $1.06$ & $-0.0118$ & $2.20$ & $0.0178$ \\
0.03  & $M_{\star}$ & $1.17$ & $-0.0007$ & $1.54$ & $0.0097$ & $2.61$ & $0.0493$ \\
      & SFR       & $0.82$ & $-0.0149$ & $1.39$ & $0.0017$ & $2.59$ & $0.0543$ \\
\hline
\end{tabular}
\end{center}
\caption{{Fit parameters of the scale-dependent bias model described by
  equation~(\ref{eqnfit}) and shown as grey lines in
  Fig.~\ref{fig:bias}. We list the values of $b_0$ and $\beta$ for
  three different samples (where $M_{200}$ stands for a selection by
  halo virial mass, $M_\star$ by stellar mass, and SFR by star
  formation rate, respectively), two different number densities
  $n$, and for three redshifts~$z$.}
\label{TabFits}
}
\end{table}

Thus far, hydrodynamical simulations of galaxy formation have been
severely challenged even by the basic task to reliably predict the
clustering length, simply due to the missing large-scale power as a
result of small simulated volumes. TNG300 is advancing the state of
the art in this regard.  At its mass resolution, no other
hydrodynamical simulation of comparable volume and with a similar
coverage of the physics exists, opening up the possibility for
quantitative studies of large-scale structure, a regime that was thus
far almost exclusively in the domain of semi-analytic models of galaxy
formation, or empirical approaches such as HOD or SHAM. This also
means that the approximate nature of these treatments can finally be
tested with full hydrodynamical simulations.

We note that the IllustrisTNG simulations not only make extant
predictions for the clustering of point objects such as galaxies and
haloes, but also for the distribution of gas and the stellar
mass, as well as, of course, for the total matter. This includes the
impact of baryonic effects on the dark matter distribution, something
that is difficult if not impossible to forecast with any degree of
reliability by (semi-)analytic models. Predicting these effects
reliably is of significant importance for the analysis of
gravitational lensing, for studying the circumgalactic medium around
galaxies, and for interpreting the population of intrahalo stars.

With TNG300, we have been able to provide accurate measurements of
the non-linear matter power spectra and correlation functions of
different mass components over a large dynamic range. We have
highlighted the strong bias of the stellar light relative to the total
matter, and the fact that its two-point correlation function is nearly
invariant in time and close to a power-law. This prediction appears to
be fairly robust to resolution and likely represents a generic feature
of $\Lambda$CDM cosmologies.

We have also shown that the galaxy distribution predicted by
IllustrisTNG clusters very similarly to observations by the Sloan
Digital Sky Survey at low redshift, both as a function of stellar mass
and galaxy colour. This is an important confirmation of the basic
validity of our hydrodynamical simulation models, and together with
the findings of our companion papers, underlines that IllustrisTNG
provides a powerful, self-consistent model for how galaxies have
emerged in the $\Lambda$CDM cosmology.

With this basic confirmation in hand, we have explored other
clustering predictions from our simulations. The observational picture
of how galaxy clustering measured in terms of clustering length and
slope of the two-point functions depends on redshift and stellar mass
has been somewhat muddled.  TNG300 makes clear statements in this
regard, showing that clustering length is a strong function of stellar
mass at all redshifts, whereas the clustering slope is not. The latter
tends to get a bit shallower towards $z\sim 1$ and hardly depends on
stellar mass over this range at all, just to become steeper again
towards high redshift and also showing again a stellar mass dependence
there.

Our analysis of galaxy and halo bias has shown that our results on the
largest scales are consistent with those obtained from simpler dark
matter only simulations. This is reassuring and largely to be
expected, given that effects of baryonic physics are restricted to
show up on intermediate and small scales. However, many observational
data sets lose significant power if their analysis is restricted to
scales that are safely unaffected by scale-dependent effects. Such
scale-dependent biases are clearly detected in our simulations,
extending out to the scales of the baryonic acoustic
oscillations. These scale-dependent biases strongly depend on the type
of tracer that is used, the sample selection criterion, the
space-density that is used, and the redshift. They originate from a
complex coupling of weakly non-linear evolution and galaxy formation
physics, something that is accounted for naturally in our
simulations. It has yet to be seen how sensitively some of these bias
predictions depend on details of the galaxy formation physics, but if
there is a chance at all to calculate them reliably, it is through
hydrodynamical simulations such as the ones discussed here.

One of the most interesting possible effects of a scale-dependent bias
is that it may impact measurements of the baryonic acoustic
oscillations based on the low-redshift galaxy distribution.  When done
in real space, the baryonic acoustic peak has been shown to be
remarkably resilient to galaxy formation physics effects
\citep[e.g.][]{Angulo2014}, but the BAO features in Fourier-space are
more drawn out in scale and thus potentially more sensitive to
distortions from a scale-dependent bias.  

While our TNG300 simulation box is just large enough to cover the
scales of the baryonic acoustic oscillations in the total matter power
spectrum, the number of available large-scale modes is unfortunately
too small to measure these weak power fluctuations directly.  However,
measurements of the scale-dependent bias of different tracers on these
scales are much less affected by cosmic variance, as this involves
dividing out the specific realisation of the total matter power
spectrum. In this way, we could demonstrate a significant
scale-dependence of the bias of different tracers over the range of
the BAOs, and also quantify the size of non-linear evolution effects
over this region. Combining both allows an estimate of the BAO
distortions in the evolved power spectrum as seen through the
tracers. We have found in this way significant shifts of the BAO peak
positions of up to 6\% in $k$, but template fitting of the expected
wiggle signal appears capable of eliminating such apparent shifts of
the acoustic scale, thereby preventing being misled in the
cosmological interpretation.

Hydrodynamical simulations of still larger volumes will be able in the
near future to substantially improve the statistics of our results on
large scales, circumventing the significant approximations involved in
other approaches to study cosmic large-scale structure and BAO
distortions from biased tracers. This offers the exciting prospect
that detailed hydrodynamical simulations of galaxy formation become an
integral and powerful part of forthcoming cosmological precision
studies.

\section*{Acknowledgements}

{We thank the anonymous referee for an insightful report.}
We thank Hellwing Wojciech for making the EAGLE power spectrum ratio
available in electronic form.  VS, RP, and RW acknowledge support
through the European Research Council under ERCStG grant
EXAGAL-308037, and would like to thank the Klaus Tschira
Foundation. The IllustrisTNG flagship simulations were run on the
HazelHen Cray XC40 supercomputer at the High-Performance Computing
Center Stuttgart (HLRS) as part of project GCS-ILLU of the Gauss
Centre for Supercomputing (GCS).  VS also acknowledges support through
subproject EXAMAG of the Priority Programme 1648 `Software for
Exascale Computing' of the German Science Foundation. MV acknowledges
support through an MIT RSC award, the support of the Alfred P. Sloan
Foundation, and support by NASA ATP grant NNX17AG29G.  JPN
acknowledges support of NSF AARF award AST-1402480.  SG and PT
acknowledge support from NASA through Hubble Fellowship grants
HST-HF2-51341.001-A and HST-HF2-51384.001-A, respectively, awarded by
the STScI, which is operated by the Association of Universities for
Research in Astronomy, Inc., for NASA, under contract NAS5-26555. The
Flatiron Institute is supported by the Simons Foundation. Ancillary
and test runs of the project were also run on the compute cluster
operated by HITS, on the Stampede supercomputer at TACC/XSEDE
(allocation AST140063), at the Hydra and Draco supercomputers at the
Max Planck Computing and Data Facility, and on the Harvard computing
facilities supported by FAS.

\bibliographystyle{mnras}
\bibliography{paper}

\label{lastpage}

\end{document}